\documentclass[preprint,12pt,3p]{elsarticle}

\usepackage{graphics}
\usepackage{graphicx}
\usepackage{subfig}
\usepackage{array}

\usepackage{amssymb}
\usepackage{amsmath}
\newcommand{\ee}{\epsilon}	

\begin{document}

\begin{frontmatter}

\title{Numerical Investigations on Quasi Steady-State Model for Voltage Stability: Limitations and Nonlinear Analysis\tnoteref{t1}}
\tnotetext[t1]{This work was supported by the Consortium for Electric Reliability
Technology Solutions provided by U.S. Department No. DE-FC26-
09NT43321.}

\author[label1]{Xiaozhe Wang\corref{cor1}}
\address[label1]{School of Electrical and Computer Engineering, Cornell University, Ithaca, NY 14853, USA}

\cortext[cor1]{Corresponding author}

\ead{xw264@cornell.edu}

\author[label1]{Hsiao-Dong Chiang}
\ead{hc63@cornell.edu}

\begin{abstract}
In this paper, several numerical examples to illustrate limitations of Quasi Steady-State (QSS) model in long-term voltage stability analysis are presented. In those cases, the QSS model provided incorrect stability assessment. Causes of failure of the QSS model are explained and analyzed in nonlinear system framework. Sufficient conditions of the QSS model for correct approximation are suggested.
\end{abstract}


\end{frontmatter}



\section{\MakeUppercase{introduction}}\label{sec1}

The complete power system dynamic models for long-term stability are large and involve different time scales. It is time-consuming and data-demanding to simulate the dynamic behaviors over long time intervals. These constraints are even more stringent in the context of on-line voltage stability assessment. People have been making efforts in long-term stability analysis from two aspects. One approach is to use automatic adjustment of step size in time domain simulation \cite{Kundur:book}\cite{de Souza:article}\cite{Jardim:article} which is from the aspect of numerical method. Another approach is to implement the quasi steady-state (QSS) model from the aspect of model approximation. The QSS model proposed in \cite{Cutsem:book}\cite{Cutsem:artical} seeks to reach a good compromise between accuracy and efficiency. By assuming that the fast variables are infinitely fast and are stable in the long-term, the QSS model replaces the differential equations of transient dynamics by their equilibrium equations \cite{Lei:article}\cite{Kumbale:article}.

Nevertheless, there are two issues associated with the QSS model. The first issue is the assumption of QSS model that the post-fault system is stable and the second issue is the assumption that the model is singularity-free. In response to these issues, a combination of the complete model and the QSS model was proposed in \cite{Cutsem:artical2} to make sure that the post-fault system is stable in short-term time scale.  In addition, a method using Newton method with optimal multiplier in \cite{Cutsem:artical3} and continuation-based QSS analysis in \cite{Wang:artical} were proposed to provide certain remedies for the singularity problem. It was conjectured that the QSS model can provide accurate approximations of the long-term stability model if the QSS model satisfy the assumptions stated above.
However, some limitations of the QSS model were indicated in \cite{Cutsem:book} saying that the QSS model can not capture the instabilities where the long-term unstable evolution triggers a short-term instability, but there was no numerical examples given. Actually, to the extent of our knowledge, there was no previous work to numerically investigate limitations of the QSS model. Additionally, there was no theoretical analysis about causes for failure of the QSS model in nonlinear system framework. In brief, the QSS model can not provide correct stability analysis of the long-term stability model consistently \cite{Wangxz:article} while little efforts have been made 
to investigate problems of the QSS model.




In this paper, several numerical examples illustrating limitations of the QSS model are presented which show that the QSS model can lead to incorrect stability assessment. In addition, causes for the failure of QSS model are explained under nonlinear system framework and also supported by numerical results. We show that the QSS model fails to provide correct approximations if either trajectory of the long-term stability model jumps outside of the stability region of the corresponding transient stability model or trajectory of the QSS model jumps away from the stable component of the constraint manifold, while these two causes for failure of the QSS model have not been mentioned in any previous work. On the other hand, explanations in nonlinear system framework are more fundamental for analysis of dynamic systems, which can lead to important steps to establish a theoretical foundation for the QSS model. 

This paper is organized as follows. Section \ref{sectioncompletemodel} reviews some relevant concepts of the long-term stability model and Section \ref{transientqss} introduces transient stability model and the QSS model which are two approximations of the long-term stability model. Nonlinear preliminaries and framework are introduced in Section \ref{preliminaries} and Section \ref{modelinnonlinear}. Section \ref{counterexample} presents three examples showing that the QSS model is stable while the long-term stability model suffers from long-term instabilities. Also, causes for the failure are explained and analyzed using nonlinear stability theories.

\section{\MakeUppercase{the long-term stability model}}

\label{sectioncompletemodel}
The long-term stability model, or interchangeably complete dynamic model, for calculating system dynamic response relative to a disturbance comprises a set of first-order differential equations and a set of algebraic equations \cite{Chiang:book}. The algebraic equations
${0}={g}({z_c,z_d,x,y})$ 
describing the electrical transmission system and the internal static behaviors of passive devices. While the transient dynamics are captured by differential equations
$\dot{{x}}={f}({z_c,z_d,x,y})$ 
which describe the internal dynamics of devices such as generators, their associated control systems, certain loads, and other dynamically modeled components. ${f}$ and ${g}$ are smooth functions, and vector ${x}$ and ${y}$ are vectors of corresponding short-term state variables and algebraic variables. If we take long-term dynamics into account, both continuous equations and discrete-time equations are needed and represented as:
$\dot{z}_{c}=\ee{h}_c({z_c,z_d,x,y})$,
${z}_d(k+1)={h}_d({z_c,z_d(k),x,y})$. 
where ${z}_c$ and ${z}_d$ are continuous and discrete long-term state variables respectively, and $1/\ee$ is the maximum time constant among devices. Usually, transient dynamics have much smaller time constants compared with those of long-term dynamics, as a result, $z_c$ and $z_d$ are also termed as slow (state) variables, and $x$ are termed as fast (state) variables. If we represent the above equations in $\tau$ time scale where $\tau=t\ee$, and denote $\prime$ as $\frac{d}{d\tau}$, then we have:
\begin{eqnarray}\label{complete}
{z}_{c}^\prime&=&{h}_c({z_c,z_d,x,y})\\
z_d(k+1)&=&h_d(z_c,z_d(k),x,y)\nonumber\\
\ee{x}^\prime&=&{f}({z_c,z_d,x,y})\nonumber\\
{0}&=&{g}({z_c,z_d,x,y})\nonumber
\end{eqnarray}

The following components need to be considered in transient time scale:

\noindent$\bullet$ Synchronous generator, and the associated excitation system;

\noindent$\bullet$ Interconnecting transmission network including static load;

\noindent$\bullet$ Induction and synchronous motor load;

\noindent$\bullet$ Other device such as HVDC converter and SVC.

In long-term stability analysis, apart from the components in transient stability analysis, it should include components such as load restoration and appropriate devices for the wide range of protection and control systems that are invoked in longer time scale. Thus the following devices need to be considered in long-term time scale \cite{Cutsem:book}\cite{Kundur:book}:

\noindent$\bullet$ Exponential and thermostatically recovery load;

\noindent$\bullet$ Turbine Governor (TG);

\noindent$\bullet$ Transformer load tap changer (LTC);

\noindent$\bullet$ OverXcitation Limiter (OXL) and armature current limiter;

\noindent$\bullet$ Shunt capacitor/reactor switching.

Note that shunt switching and LTC changing are typical events that result in discrete dynamics captured by ${z}_d(k+1)={h}_d({z_c,z_d(k),x,y})$.

Power system dynamic behaviors after a contingency are fairly complex and it requires a lot of computational efforts to do time domain simulation of the long-term stability model over long time intervals. A natural idea is to decompose the long-term stability model in order to reach a good approximation of the long-term stability model in different time scales. Hence, the transient stability model in short term and the QSS model in long term were proposed and both models are going to be briefly introduced in the next Section.
\section{\MakeUppercase{transient stability model and the qss model}}\label{transientqss}
Conventional practice in power system analysis has been to use the simplest acceptable system, which captures the essence of the phenomenon under study \cite{Chiang:book}. 
Transient model:
\begin{eqnarray}
\dot{{x}}&=&{f}({z_c,z_d,x,y})\\
 {0}&=&{g}({z_c,z_d,x,y})\nonumber
 \end{eqnarray}
is obtained by assuming that slow variables ${z}_c$ and ${z}_d$ are constant parameters.
Transient model has been widely used in industry and research for transient stability analysis. Detailed model of generators, exciters as well as motors, HVDC converters and SVC are included, while dynamics of the other components are not included in the study. It is because exponential or thermostatically recovery loads, and turbine governors have very slow responses from the viewpoint of transient stability \cite{Kundur:book}, while LTC and OXL even don't start to work in transient time scale. Hence, the model used in transient stability analysis regard slow variables as constants.

While in the QSS model, the dynamic behavior of fast variables can be considered as instantaneously fast in long-term time scale and can be replaced by its equilibrium equations $0={f}({z_c,z_d,x,y})$. 

Additionally, since transitions of $z_d$ depend on system variables, thus $z_d$ change values from $z_d(k-1)$ to $z_d(k)$ at distinct times $\tau_k$ where $k=1,2,3,...N$, otherwise, these variables remain constants. Therefore the long-term stability model (\ref{complete}) can be regarded as two decoupled systems (\ref{couple1}) and (\ref{couple2}) shown below. When $z_d$ change by (\ref{couple1}):
\begin{equation}
z_d(k+1)=h_d(z_c,z_d(k),x,y) \label{couple1}
\end{equation}
system (\ref{couple2}) works with freezing parameters $z_d$:
\begin{eqnarray}\label{couple2}
{z}_{c}^\prime&=&{h}_c({z_c,z_d(k+1),x,y})\\
\ee{x}^\prime&=&{f}({z_c,z_d(k+1),x,y})\nonumber\\
{0}&=&{g}({z_c,z_d(k+1),x,y})\nonumber
\end{eqnarray}

Similarly, the QSS model:
\begin{eqnarray}\label{QSS}
{z}_{c}^\prime&=&{h}_c({z_c,z_d,x,y})\\
z_d(k+1)&=&h_d(z_c,z_d(k),x,y)\nonumber\\
{0}&=&{f}({z_c,z_d,x,y})\nonumber\\
{0}&=&{g}({z_c,z_d,x,y})\nonumber
\end{eqnarray}
can also be decoupled as:
\begin{equation}
z_d(k+1)=h_d(z_c,z_d(k),x,y) \label{coupleqss1}
\end{equation}
and
\begin{eqnarray}\label{coupleqss2}
{z}_{c}^\prime&=&{h}_c({z_c,z_d(k+1),x,y})\\
{0}&=&{f}({z_c,z_d(k+1),x,y})\nonumber\\
{0}&=&{g}({z_c,z_d(k+1),x,y})\nonumber
\end{eqnarray}


Generally the QSS model is able to provide correct approximations of the long-term stability model with correct stability assessment, while the QSS model takes much less time if larger time steps or adaptive time steps are implemented. Also, compared with the long-term stability model, the Jacobian matrix of QSS model can be updated only following discrete events unless slow convergence rate is observed \cite{Cutsem:book}. However as stated at the beginning, the QSS model can not capture the dynamic behavior of the long-term stability model and fail to provide correct approximations of the long-term stability model in some cases. And little attention has been paid to this severe problem. The failure of the QSS model is going to be further analyzed in Section \ref{counterexample}.

\begin{figure}[!ht]
\centering
\begin{minipage}[t]{0.5\linewidth}
\includegraphics[width=2.5in ,keepaspectratio=true,angle=0]{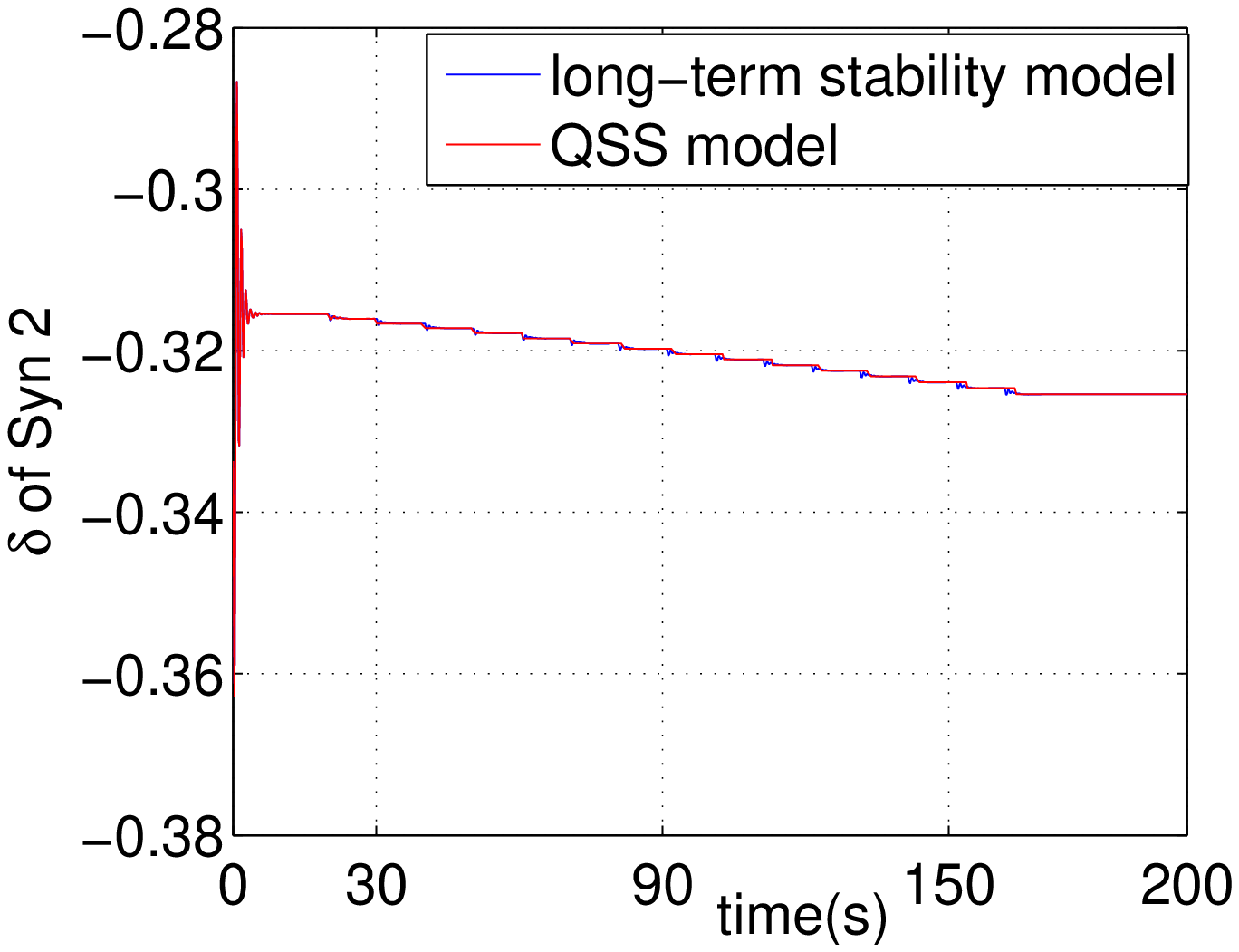}
\end{minipage}%
\begin{minipage}[t]{0.5\linewidth}
\includegraphics[width=2.5in ,keepaspectratio=true,angle=0]{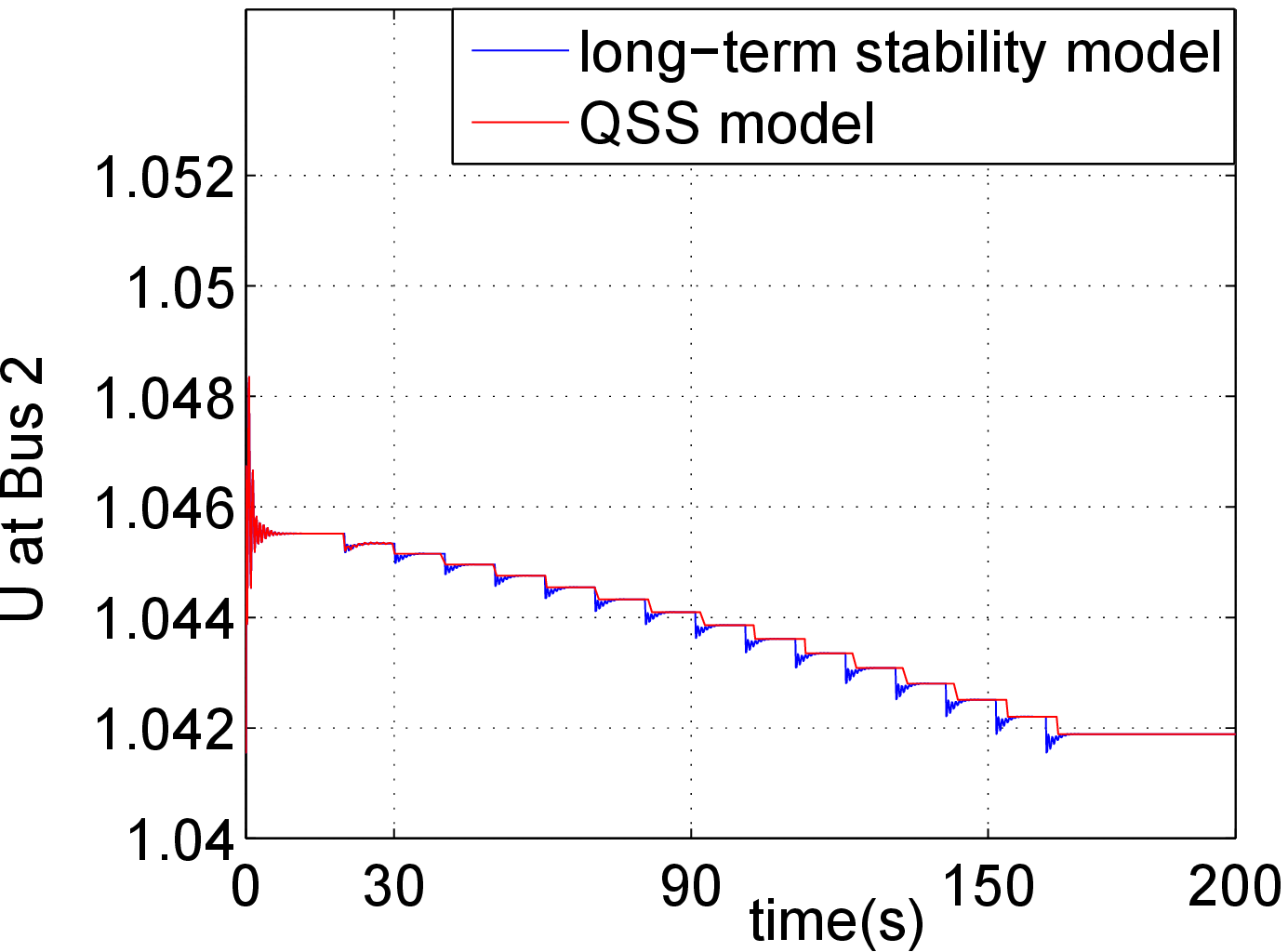}
\end{minipage}
\caption{The trajectory comparisons of the long-term stability model and the QSS model for different variables. The trajectory of the long-term stability model followed that of the QSS model until both of them converged to the same long-term SEP.}\label{my14completeqss}
\end{figure}

At the end of this section, a numerical example based on the IEEE 14-bus system is presented in Fig. \ref{my14completeqss}. In this numerical example, the QSS model successfully provided accurate approximations of the long-term stability model, and both of them finally settled down to the same long-term stable equilibrium point (SEP). Thus the long-term stability of the QSS model implied the stability of the long-term stability model in this case.

\section{\MakeUppercase{nonlinear system preliminaries}}\label{preliminaries}
In this section, some relevant stability concepts from the nonlinear stability theories are briefly introduced. Knowledge of the stability region is essential in analyzing the QSS model in long-term stability analysis.

\subsection{{Stability of Equilibrium Point and Stability Region}}
We consider the following autonomous nonlinear dynamical system:
\begin{equation}\label{general autonomous}
\dot{x}=f(x), \quad x\in\Re^n
\end{equation}
where $f:\Re^n\rightarrow\Re^n$ satisfies a sufficient condition for the existence and uniqueness of a solution. The solution of (\ref{general autonomous}) starting at initial state $x$ at time $t=0$ is called the system trajectory and is denoted as $\phi(t,x)$. $\bar{x}\in\Re^n$ is said to be an equilibrium point of (\ref{general autonomous}) if $f(\bar{x})=0$. The definition of asymptotic stability is given as below \cite{Chiang:book}:

\noindent\textit{Definition 1: Asymptotic Stability}

An equilibrium point $\bar{x}\in\Re^n$ of (\ref{general autonomous}) is said to be asymptotically stable if, for each open neighborhood $U$ of $\bar{x}\in\Re^n$, the followings are true: (i) $\phi(t,x)\in{U}$ for all $t>0$; (ii) $\lim_{t \to \infty}\parallel\phi(t,x)-\bar{x}\parallel=0$.

Without confusion, we use stable equilibrium point (SEP) instead of asymptotically stable equilibrium point in this paper. An equilibrium point is \textit{hyperbolic} if the corresponding Jacobian matrix has no eigenvalues with zero real parts. And a hyperbolic equilibrium point $\bar{x}$ is a \textit{type-k equilibrium point} if there exist $k$ eigenvalues of $D_xf(\bar{x})$ with positive real parts.

The stability region of a SEP $x_s$ is the set of all points $x$ such that $\lim_{t \to \infty}\phi(t,x)\rightarrow{x_s}$. In other words, the \textit{stability region} is defined as: $A(x_s):=\{x\in\Re^n:\lim_{t \to \infty}\phi(t,x)=x_s\}$. From a topological point of view, the stability region is an open invariant and connected set. Every trajectory in a stability region lies entirely in the stability region and the dimension of the stability region is $n$.
\subsection{{Singular Perturbed System}}
Consider the following general singular perturbed model:
\begin{eqnarray}\label{singular perturb}
\Sigma_\ee:\dot{z}&=&f(z,x)\qquad z\in\Re^n\\
\ee\dot{x}&=&g(z,x)\qquad x\in\Re^m\nonumber
\end{eqnarray}
where $\ee$ is a small positive parameter. $z$ is a vector of slow variables while $x$ is a vector of fast variables. Let $\phi_\ee(t,z_0,x_0)$ denotes the trajectory of model (\ref{singular perturb}) starting at $(z_0,x_0)$ and $E$ denotes the set of equilibrium points of it, i.e. $E=\{(z,x)\in\Re^n\times\Re^m: f(z,x)=0, g(z,x)=0\}$. If $(z_s,x_s)$ is a SEP of model (\ref{singular perturb}), then the stability region of $(z_s,x_s)$ is defined as:
\begin{equation}
A_\ee(z_s,x_s):=\{(z,x)\in\Re^n\times\Re^m: \phi_\ee(t,z_0,x_0)\rightarrow
(z_s,x_s)\mbox{ as }t\rightarrow\infty\}\nonumber
\end{equation}
The slow model (quasi steady-state model) is obtained by setting $\ee=0$ in (\ref{singular perturb}):
\begin{eqnarray}\label{slow}
\Sigma_0:\dot{z}&=&f(z,x)\qquad z\in\Re^n\\
0&=&g(z,x)\qquad x\in\Re^m\nonumber
\end{eqnarray}
The algebraic equation $0=g(z,x)$ constraints the slow dynamics to the following set which is termed as \textit{constraint manifold}:
\begin{equation}
\Gamma:=\{(z,x)\in\Re^n\times\Re^m:g(z,x)=0\}
\end{equation}
The trajectory of model (\ref{slow}) starting at $z_0$ is denoted by $\phi_0(t,z_0,x_0)$ and the stability region is
\begin{equation}
A_0(z_s,x_s):=\{(z,x)\in\Gamma: \phi_0(t,z_0)\rightarrow(z_s,x_s)\mbox{ as}\rightarrow\infty\}\nonumber
\end{equation}

The \textit{singular points} of model (\ref{slow}) or \textit{singularity $S$} is defined as:
\begin{equation}
S:=\{(z,x)\in\Gamma:\mbox{ det}(D_xg)(z,x)=0\}
\end{equation}
Singular points can drastically influence the trajectories of the differential-algebraic equation (DAE) system since the validity of the model (\ref{slow}) typically breaks down at singular points. Typically, the singular set $S$ is a stratified set of maximal dimension $n-1$ embedded in $\Gamma$ and $\Gamma$ is separated by $S$ into open regions \cite{Veukatasubramanian:article}. 

\noindent \textit{Definition 2: Type of Constraint Manifold}

The connected set $\Gamma_i\subset\Gamma$ is a \textit{type-k component of $\Gamma$} if the matrix $D_xg$, evaluated at every point of $\Gamma_i$, has $k$ eigenvalues that have positive real parts. If all the eigenvalues of $D_xg$ calculated at points of $\Gamma_i$ have a negative real part, then we call $\Gamma_i$ a \textit{stable component of $\Gamma$}; otherwise, it's an \textit{unstable component of $\Gamma$}.

Next we proceed to define the fast model (boundary layer model). Define the fast time scale $\sigma=t/\ee$. In this time scale, model (\ref{singular perturb}) takes the form:
\begin{eqnarray}\label{singular perturb in sigma}
\Pi_\ee:\frac{dz}{d\sigma}&=&{\ee}f(z,x)\qquad z\in\Re^n\\
\frac{dx}{d\sigma}&=&g(z,x)\qquad x\in\Re^m\nonumber
\end{eqnarray}
Then the fast model is defined as:
\begin{equation}\label{BLS}
\Pi_f:\frac{dx}{d\sigma}=g(z,x)
\end{equation}
where $z$ is frozen and treated as a parameter. The constraint manifold $\Gamma$ is a set of equilibriums of models (\ref{BLS}). For each fixed $z$, a fast dynamical model (\ref{BLS}) is defined.

A comprehensive theory of stability regions can be found in \cite{Chiang:article1988}-\cite{Alberto:article}.

\section{\MakeUppercase{models in nonlinear framework }}\label{modelinnonlinear}


Assuming $(z_{cls},z_{dls},x_{ls},y_{ls})$ is a long-term SEP of both the long-term stability model (\ref{complete}) and the QSS model (\ref{QSS}) in the study region $U$ starting from $(z_{c0},z_{d0},x_0^l,y_0^l)$ and $(z_{c0},z_{d0},x_0^q,y_0^q)$ respectively, and $\phi_l(\tau,z_{c0},z_{d0},x_0^l,y_0^l)$ denotes the trajectory of the long-term stability model (\ref{complete}) and $\phi_q(\tau,z_{c0},z_{d0},x_0^q,y_0^q)$ denotes the trajectory of QSS model (\ref{QSS}). Then stability region of the long-term model (\ref{complete}) is:
\begin{equation}
A_l(z_{cls},z_{dls},x_{ls},y_{ls}):=\{(z_c,z_d,x,y)\in{U}:\phi_l(\tau,z_{c0},z_{d0},x_0^l,y_0^l)\rightarrow(z_{cls},z_{dls},x_{ls},y_{ls})\mbox{ as $\tau$}\rightarrow+\infty\}\nonumber
\end{equation}
For the QSS model (\ref{QSS}), since its dynamics are constrained to the manifold:
$\Gamma:=\{(z_c,z_d,x,y)\in{U}:{f}({z_c,z_d,x,y})=0, {g}({z_c,z_d,x,y})=0\}$.
Then the stability region of the QSS model (\ref{QSS}) is
\begin{equation}
A_q(z_{cls},z_{dls},x_{ls},y_{ls}):=\{(z_c,z_d,x,y)\in{\Gamma}:\phi_q(\tau,z_{c0},z_{d0},x_0^q,y_0^q)\rightarrow(z_{cls},z_{dls},x_{ls},y_{ls})\mbox{ as $\tau$}\rightarrow+\infty\}\nonumber
\end{equation}
Note that the constraint manifold $\Gamma$ may not be smooth due to the discrete behavior of $z_d$.

The singular points of constraint manifold $\Gamma$ is:
\begin{equation}
S:=\{(z_c,z_d,x,y)\in\Gamma:\mbox{det}\left[\begin{array}{cc}D_xf&D_yf\\ D_xg&D_yg\end{array}\right]=0\}
\end{equation}
And stable component of $\Gamma$  
is defined as:
\begin{equation}
\Gamma_0=\{(z_c,z_d,x,y)\in\Gamma: \mbox{all eigenvalues }\lambda\mbox{ of }\left[\begin{array}{cc}D_xf&D_yf\\ D_xg&D_yg\end{array}\right]
\mbox{ satisfy Re}(\lambda)<0\}
\end{equation}

For each fixed $z_c^\star$ and $z_d(k)$, given a point $(z_c^\star,z_d(k),x,y)$ on $\Gamma$, the corresponding transient stability model (i.e. the fast model) is defined as:
\begin{eqnarray}\label{transient}
\dot{x}&=&{f}({z_c^{\star},z_d(k),x,y})\\
{0}&=&{g}({z_c^{\star},z_d(k),x,y})\nonumber
\end{eqnarray}
If $(z_c,z_d(k),x,y)\not\in S$, then
$(z_c,z_d(k),x_{ts},y_{ts})$ is an equilibrium point of (\ref{transient}), where
\begin{equation}
\left(\begin{array}{cc}x_{ts}\\y_{ts}\end{array}\right)=\left(\begin{array}{cc}l_1(z_c,z_d(k))\\l_2(z_c,z_d(k))\end{array}\right)=l(z_c,z_d(k))\nonumber
\end{equation}
$(z_c^\star,z_d(k),x_{ts},y_{ts})$ is termed as transient SEP whose stability region is represented as:
\begin{eqnarray*}\label{transientsep}
&&A_t(z_c^{\star},z_d(k),x_{ts},y_{ts}):=\{(x,y)\in U,z_c=z_c^{\star},z_d=z_d(k):\nonumber\\
&&\phi_t(t,z_c^{\star},z_d(k),x,y)\rightarrow(z_c^{\star},z_d(k),x_{ts},y_{ts})\mbox{as t}\rightarrow+\infty\}\nonumber\\
\end{eqnarray*}
where $\phi_t(t,z_c^{\star},z_d(k),x,y)$ denotes trajectory of the transient stability model (\ref{transient}).

Assuming that $D_y g$ is nonsingular, then the transient stability model (\ref{transient}) can be represented as
\begin{equation}
\dot{x}={f}({z_c^{\star},z_d(k),x,j(z_c^{\star},z_d(k),x)})
\end{equation}
where $y=j(z_c^{\star},z_d(k),x)$ is an isolated root of ${g}({z_c^{\star},z_d(k),x,y})=0$.
And we can define the following $\Gamma_s$ which is a subset of $\Gamma_0$:
\begin{eqnarray*}\label{gammas}
&&\Gamma_s=\{(z_c,z_d,x,y)\in\Gamma: \mbox{all eigenvalues $\lambda$ of }(D_x f\nonumber-D_y f{D_y g}^{-1}D_x g)\nonumber\\
&&\mbox{ satisfy Re}(\lambda)<0, \mbox{ and }D_y g\mbox{ is nonsingular}\}\nonumber\\
\end{eqnarray*}
such that each point on $\Gamma_s$ is a SEP of the corresponding transient stability model defined in (\ref{transient}) for each fixed $z_c^{\star}$ and $z_d(k)$.

\begin{figure}[!ht]
\centering
\includegraphics[width=4in,keepaspectratio=true,angle=0]{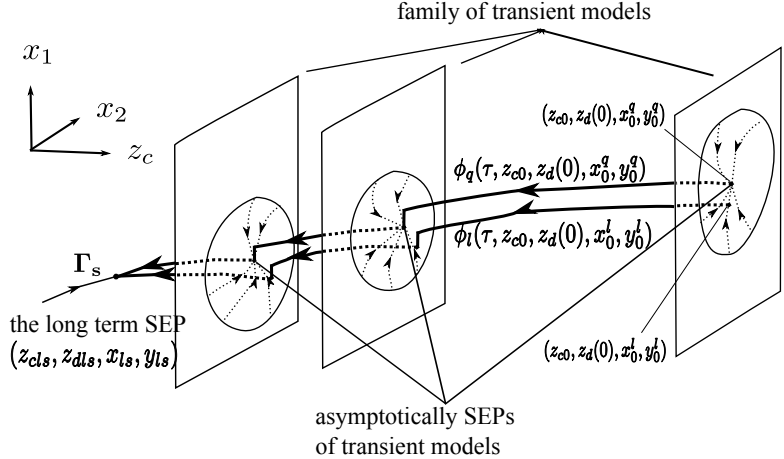}\caption{The dynamic relation between different models. $\phi_q(\tau,z_{c0},z_{d0},x_0^q,y_0^q)$ is constrained on the slow manifold which is a subset of $\Gamma_s$ and each point of $\phi_l(\tau,z_{c0},z_{d0},x_0^l,y_0^l)$ locates inside the stability region of the corresponding transient stability model, finally both $\phi_l(\tau,z_{c0},z_{d0},x_0^l,y_0^l)$ and $\phi_q(\tau,z_{c0},z_{d0},x_0^q,y_0^q)$ converge to the same long-term SEP $(z_{cls},z_{dls},x_{ls},y_{ls})$. }\label{QSSstable}
\end{figure}

$\phi_q(\tau,z_{c0},z_{d0},x_0^q,y_0^q)$ is constrained on a certain set of $\Gamma$, for convenience, we term this set to be the slow manifold of the QSS model. Similarly, $\phi_l(\tau,z_{c0},z_{d0},x_0^l,y_0^l)$ stays close to a set of $\Gamma$ which is called the slow manifold of the long-term stability model. Generally, the slow manifold of the QSS model is a subset of $\Gamma_s$ such that each point of $\phi_q(\tau,z_{c0},z_{d0},x_0^q,y_0^q)$ is a SEP of the corresponding transient stability model. And $\phi_l(\tau,z_{c0},z_{d0},x_0^l,y_0^l)$ always stays close to the slow manifold of the QSS model, or more rigorously speaking, each point $(z_c^\star,z_d(k),x,y)$ of $\phi_l(\tau,z_{c0},z_{d0},x_0^l,y_0^l)$ is in the stability region $A_t(z_c^{\star},z_d(k),x_{ts},y_{ts})$ of the corresponding transient stability model. Finally both $\phi_q(\tau,z_{c0},z_{d0},x_0^q,y_0^q)$ and $\phi_l(\tau,z_{c0},z_{d0},x_0^l,y_0^l)$ converge to the same long-term SEP. The dynamic relation between the long-term stability model and the QSS model is illustrated in Fig. \ref{QSSstable}.


\section{\MakeUppercase{Numerical Investigations}}\label{counterexample}

In this section, several numerical examples in which the QSS model failed to provide correct approximations of the long-term stability model are presented, and all simulations were done using PSAT 2.1.6 \cite{Milano:article}. In general, there are two causes for the failure. Firstly, the trajectory of the long-term stability model gets outside of the stability region of the corresponding transient stability model. 
Secondly, the trajectory of the QSS model jumps away from the stable component of the constraint manifold triggered by the evolution of discrete variables.

\subsection{{Numerical Example I: 9-bus power system}}\label{9bus}
\begin{figure}[!ht]
\centering
\begin{minipage}[t]{0.5\linewidth}
\includegraphics[width=2.5in ,keepaspectratio=true,angle=0]{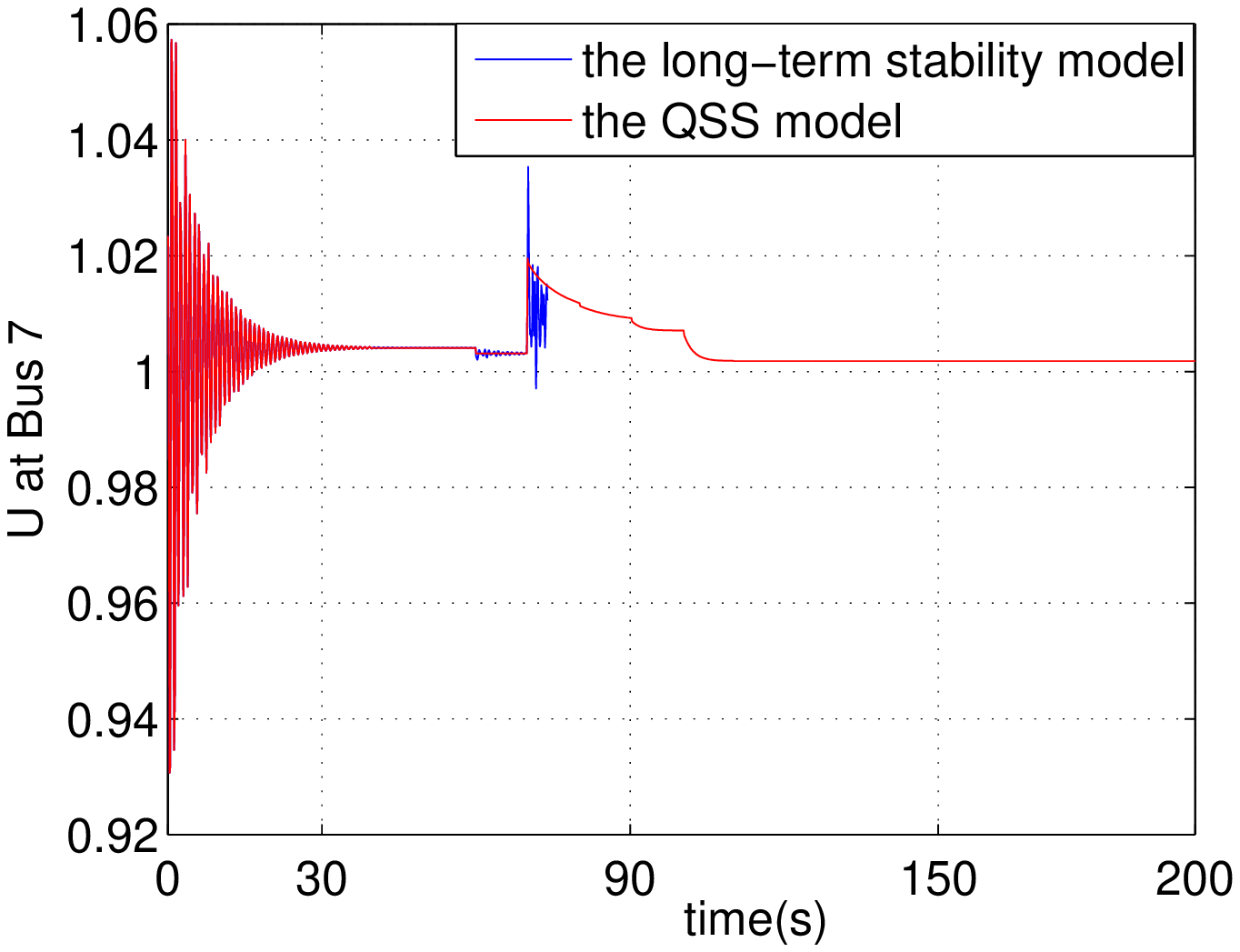}
\end{minipage}%
\begin{minipage}[t]{0.5\linewidth}
\includegraphics[width=2.5in ,keepaspectratio=true,angle=0]{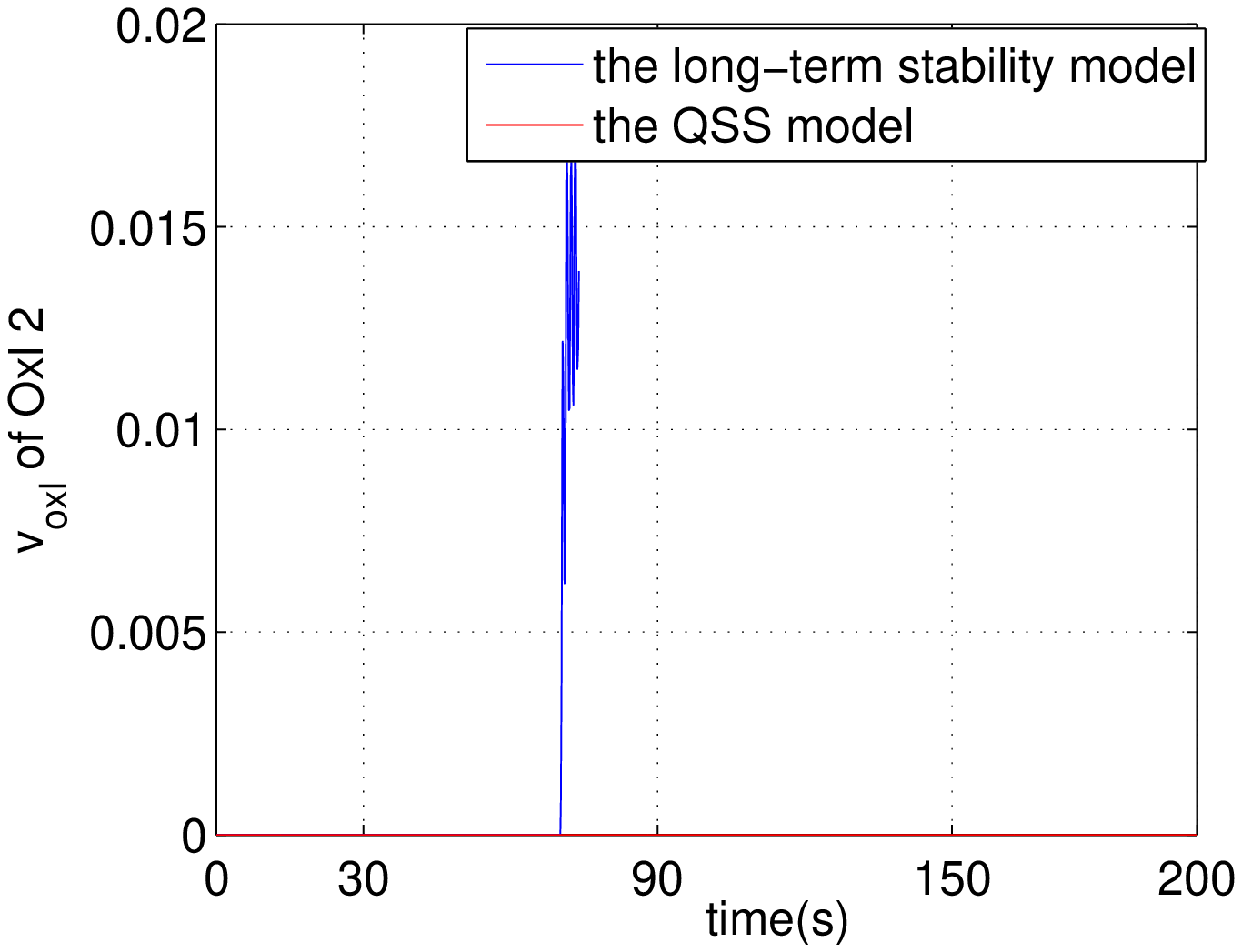}
\end{minipage}
\caption{The trajectory comparisons of the long-term stability model and the QSS model. In this case, the QSS model converged to a long-term SEP while the long-term stability model stopped at 73.9351s due to instabilities caused by wild oscillation of fast variables.} 
\label{my14completeqss_try}
\label{my9completeqss}
\end{figure}

A 9-bus power system is presented in which the QSS model failed to provide correct approximations of the long-term stability model and yielded incorrect stability assessment. In this system, each generator was controlled by AVR, TG and OXL. There was an exponential recovery load at Bus 5. The system was highly stressed after the fault and the voltage stability margin was less than 40\%. Trajectory comparisons between the long-term stability model and the QSS model are shown in Fig. \ref{my9completeqss}. In order to make sure that the post-fault system was stable in short-term time scale, the QSS model was implemented starting from 40s before which the long-term stability model was used. The QSS performed well from 40s to 60s until LTCs and OXLs started to work. When $z_d$ firstly changed at 60s, trajectories of the long-term stability model and the QSS model stayed close to each other and the QSS model still worked properly. While $z_d$ changed again at 70s, the long-term stability model was no longer stable which can be seen from the wild oscillations of variables, and the simulation stopped at 73.9351s. Nevertheless, the QSS model continued on until it converged to a long-term SEP. Since voltages of all buses were remained between 0.95 p.u and 1.05 p.u, and all generators were synchronized, the results of the QSS model indicated that the post-fault system was stable in long-term stability analysis which contradicted the results of the long-term stability model.

\begin{figure}[!ht]
\centering
\includegraphics[width=4in ,keepaspectratio=true,angle=0]{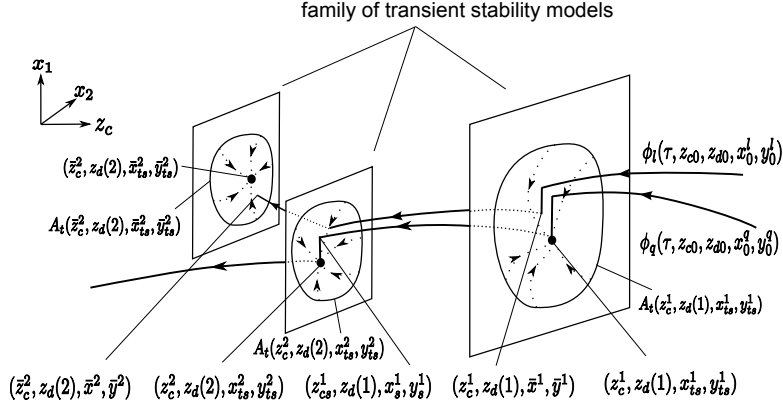}\caption{The dynamic relation between the long-term stability model and the QSS model. $\phi_l(\tau,z_{c0},z_{d0},x_0^l,y_0^l)$ jumped far away from the the slow manifold of the QSS model when $z_d$ changed to $z_d(2)$ such that slow manifolds of the QSS model and the long-term stability model were separated.}\label{my9case2illustration}
\end{figure}

The main reason for this failure was that when discrete variables changed from $z_d(1)$ to $z_d(2)$, the distance between $z_c$ of two models sharply increased such that $\phi_l(\tau,z_{c0},z_{d0},x_0^l,y_0^l)$ jumped away from the slow manifold of the QSS model immediately. An illustration of dynamic relation between $\phi_q(\tau,z_{c0},z_{d0},x_0^q,y_0^q)$ and $\phi_l(\tau,z_{c0},z_{d0},x_0^l,y_0^l)$ is shown in Fig. \ref{my9case2illustration}. Assuming after $z_d$ changed from $z_d(0)$ to $z_d(1)$ at 60s, $\phi_q(\tau,z_{c0},z_{d0},x_0^q,y_0^q)$ jumped to $(z_c^1,z_d(1),x_{ts}^1,y_{ts}^1)$, and $\phi_l(\tau,z_{c0},z_{d0},x_0^l,y_0^l)$ jumped to $(z_c^1,z_d(1),\bar{x}^1,\bar{y}^1)$,  both the long-term stability model and the QSS model almost settled down to $(z_{cs}^1,z_d(1),x_{s}^1,y_{s}^1)$ by 70s, then when $z_d$ changed from $z_d(1)$ to $z_d(2)$ at 70s, $\phi_q(\tau,z_{c0},z_{d0},x_0^q,y_0^q)$ jumped to $(z_c^2,z_d(2),x_{ts}^2,y_{ts}^2)$, while $\phi_l(\tau,z_{c0},z_{d0},x_0^l,y_0^l)$ jumped to $(\bar{z}_c^2,z_d(2),\bar{x}^2,\bar{y}^2)$, and $||z_c^2-\bar{z}_c^2||$ increased immediately such that slow manifolds of the long-term stability model and the QSS model got separated.  Note that $(\bar{z}_c^2,z_d(2),\bar{x}^2,\bar{y}^2)$ was the initial point of the long-term stability model:
\begin{eqnarray}\label{complete 3}
{z}_{c}^\prime&=&{h}_c({z_c,z_d(2),x,y}),\hspace{0.36in} z_c(\tau_0)=\bar{z}_c^2\\
\ee{x}^\prime&=&{f}({z_c,z_d(2),x,y}),\qquad\quad x(\tau_0)=\bar{x}^2\nonumber,\\
{0}&=&{g}({z_c,z_d(2),x,y})\nonumber
\end{eqnarray}
and this initial point was outside of the stability region $A_t(z_c^2,z_d(2),x_{ts}^2,y_{ts}^2)$ of the transient stability model:
\begin{eqnarray}\label{transientqss 3}
\dot{x}&=&{f}({z_c^2,z_d(2),x,y}),\qquad\quad x(t_0)=x_{ts}^2 \\
{0}&=&{g}({z_c^2,z_d(2),x,y})\nonumber
\end{eqnarray}


\begin{figure}[!ht]%
\centering
\subfloat[]{\label{my14sepchangetransient_61}\includegraphics[width=2.5in ,keepaspectratio=true,angle=0]{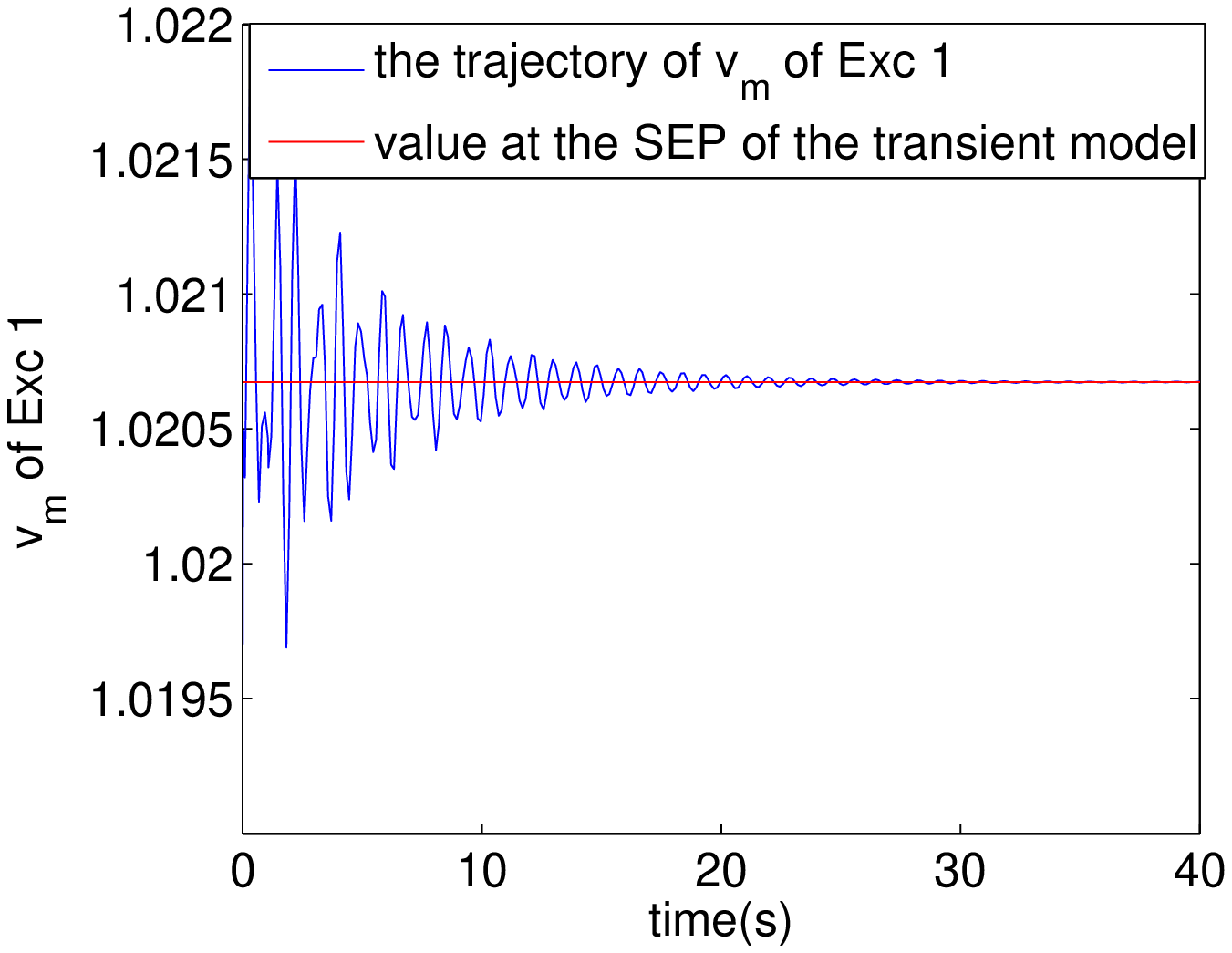}}
\subfloat[]{\label{my14sepchangetransient_71}\includegraphics[width=2.5in ,keepaspectratio=true,angle=0]{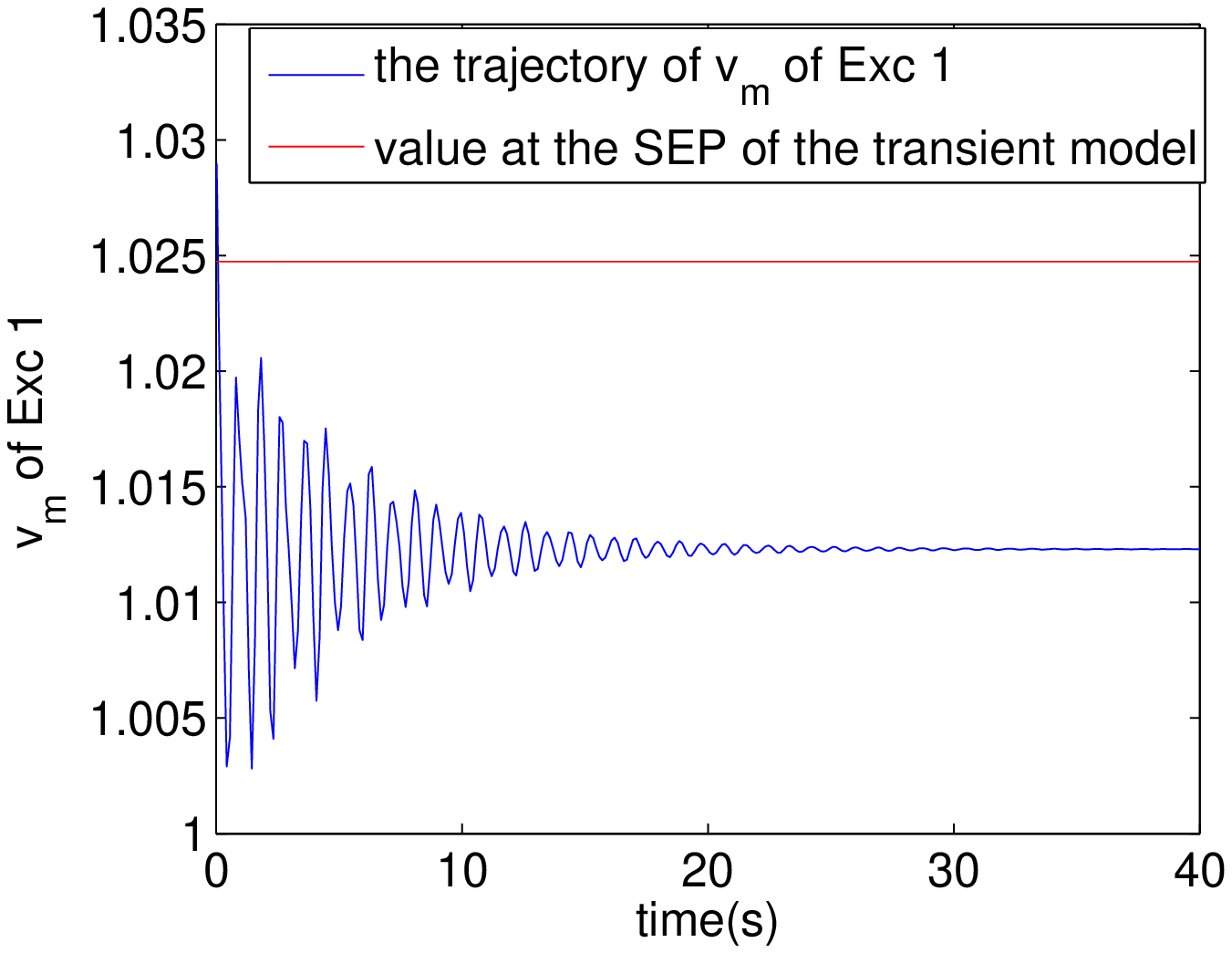}}
\caption{(a).Trajectory of the transient stability model when LTCs changed at 60s. The trajectories starting from $(z_c^1,z_d(1),\bar{x}^1,\bar{y}^1)$ converged to the SEP of the transient stability model. (b). Trajectory of the transient stability model when LTCs changed the second time at 70s. The trajectories starting from $(z_c^2,z_d(2),\bar{x}^2,\bar{y}^2)$ didn't converge to the SEP of the transient stability model. }
\end{figure}

Additionally, trajectory of a fast variable of the transient stability model:
\begin{eqnarray}\label{transientqss 3}
\dot{x}&=&{f}({z_c^1,z_d(1),x,y}),\qquad\quad x(t_0)=\bar{x}^1 \\
{0}&=&{g}({z_c^1,z_d(1),x,y})\nonumber
\end{eqnarray}
is shown in Fig. \ref{my14sepchangetransient_61}, and it can be seen when $z_d$ changed the first time, $(z_c^1,z_d(1),\bar{x}^1,\bar{y}^1)$ on $\phi_l(\tau,z_{c0},z_{d0},x_0^l,y_0^l)$ was inside the stability region $A_t(z_c^1,z_d(1),x_{ts}^1,y_{ts}^1)$ of the corresponding transient stability model.
However, when $z_d$ changed from $z_d(1)$ and $z_d(2)$, the point $(z_c^2,z_d(2),\bar{x}^2,\bar{y}^2)$ was no longer inside the stability region $A_t(z_c^2,z_d(2),x_{ts}^2,y_{ts}^2)$ of the corresponding transient stability model as shown in Fig. \ref{my14sepchangetransient_71}.


Therefore, the key reason for this failure was that slow manifolds of two models got separated and the trajectory of the long-term stability model jumped outside the stability region of the corresponding transient stability model.


\subsection{{Numerical Example II: 14-bus system}}\label{14bus}
Next, another example of the QSS model is presented. In this 14-bus system, each generator was controlled by AVR and OXL, and there were three exponential recovery load at Bus 9, Bus 10 and Bus 14 respectively. Besides, there were two turbine governors at Generator 1 and Generator 3. Actually, the system had the same network topology and models as that of Numerical Example II in \cite{Wangxz:article}, but with different loading conditions and parameters. 
Since here comes another example where the QSS model failed, it indicates that this situation is not rare and definitely needs more attentions. %

\begin{figure}[!ht]
\centering
\begin{minipage}[t]{0.5\linewidth}
\includegraphics[width=2.5in ,keepaspectratio=true,angle=0]{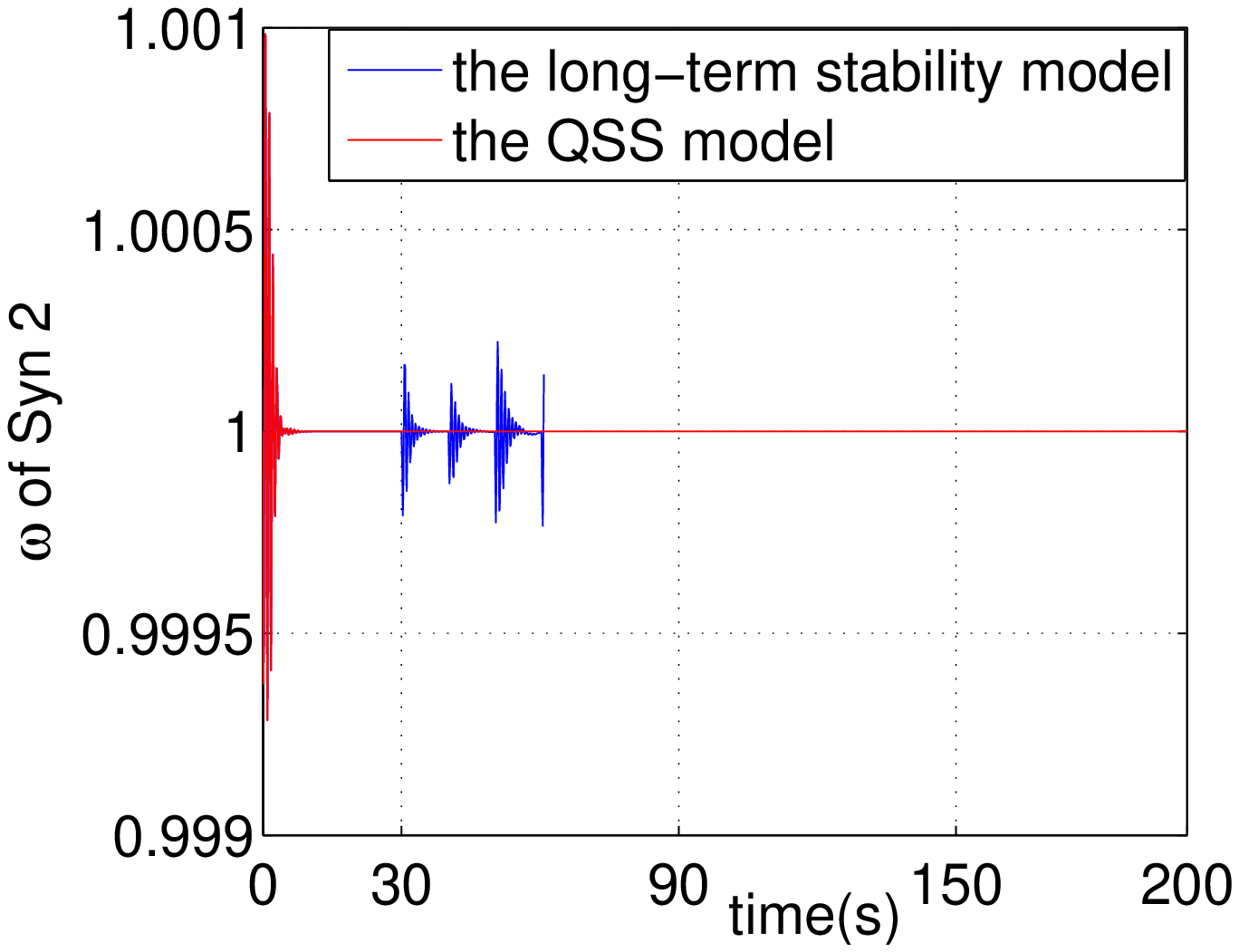}
\end{minipage}%
\begin{minipage}[t]{0.5\linewidth}
\includegraphics[width=2.5in ,keepaspectratio=true,angle=0]{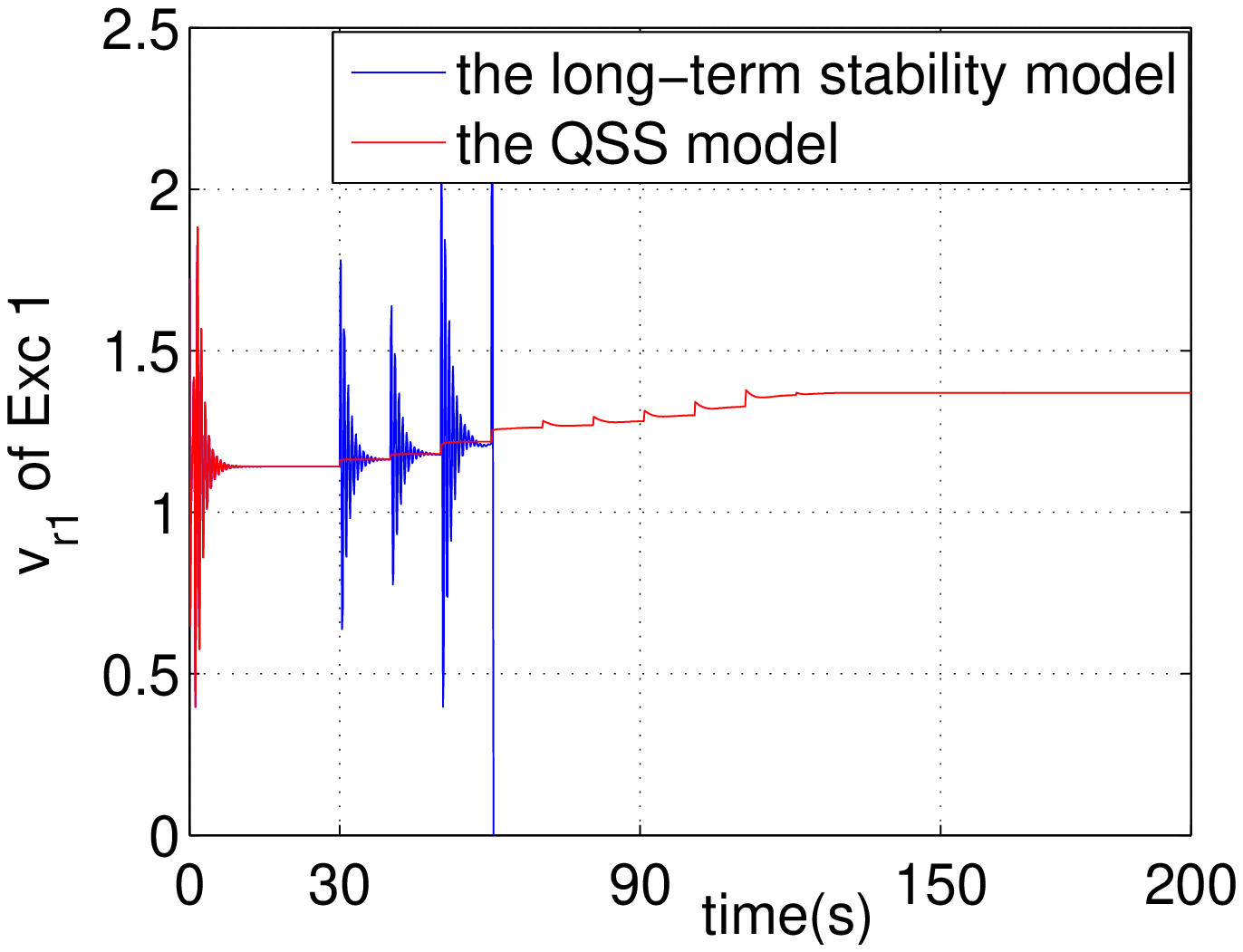}
\end{minipage}
\caption{The trajectory comparisons of the long-term stability model and the QSS model. In this case, the QSS model converged to a long-term SEP while the long-term stability model stopped at 60.7692s due to instabilities caused by wild oscillation of transient variables.} 
\label{my14completeqss_sepchange}
\end{figure}

Trajectories of the long-term stability model and the QSS model are shown in Fig. \ref{my14completeqss_sepchange}. The post-fault system was stable in short-term time scale, thus the QSS model started to work at 30s and LTCs as well as OXLs started to work at the same time. The system was highly stressed after the fault and the voltage stability margin was less than 50\%. At the beginning, trajectories of both the QSS model and the long-term stability model stayed close to each other, while the oscillations of variables in the long-term stability model became severer, and finally the simulation of the long-term stability model stopped at 60.7692s. On the other hand, the QSS model didn't capture those dynamics and converged to a long-term SEP by the end of 200s. Since voltages of all buses were remained in a reasonable range and all generators were synchronized, the results of the QSS model indicated that the post-fault system was stable in long-term stability analysis. Thus the QSS model failed to give correct approximations of the long-term stability model with incorrect stability assessment again.

\begin{figure}[!ht]
\centering
\includegraphics[width=4in ,keepaspectratio=true,angle=0]{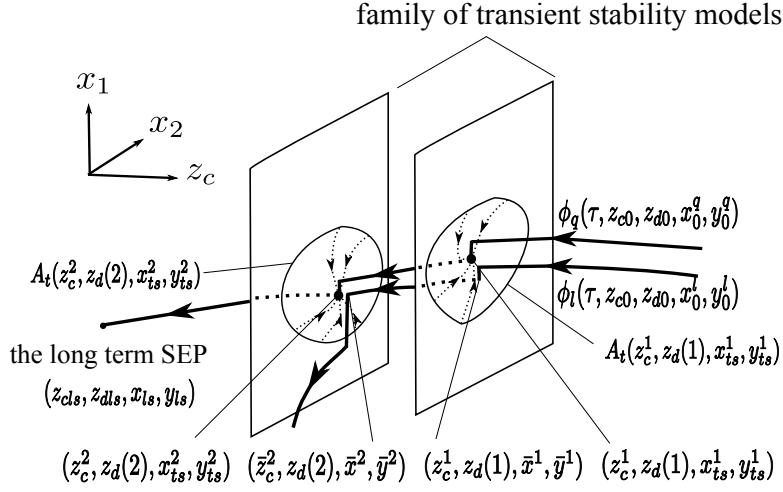}\caption{The dynamic relation between the long-term stability model and the QSS model. $\phi_l(\tau,z_{c0},z_{d0},x_0^l,y_0^l)$ jumped far away from the the slow manifold of the QSS model when $z_d$ firstly changed to $z_d(2)$ such that slow manifolds of the QSS model and the long-term stability model were separated.}\label{my14illustration}
\end{figure}

The reason for failure was that the initial point of the long-term stability model fixed at $z_d(2)$ was outside of the stability region while the difference between this example and the last one is that distance between $z_c$ of two models didn't increase immediately after $z_d$ changed. However, similarly to the last example, the slow manifolds of two models were separated and the trajectories of two models moved far away from each other afterwards. There are more details as follows. An illustration of dynamic relation between $\phi_l(\tau,z_{c0},z_{d0},x_0^l,y_0^l)$ and $\phi_q(\tau,z_{c0},z_{d0},x_0^q,y_0^q)$ is shown in Fig. \ref{my14illustration}.

\begin{figure}[!ht]%
\centering
\subfloat[]{\label{my14sepchange_31}\includegraphics[width=2.5in ,keepaspectratio=true,angle=0]{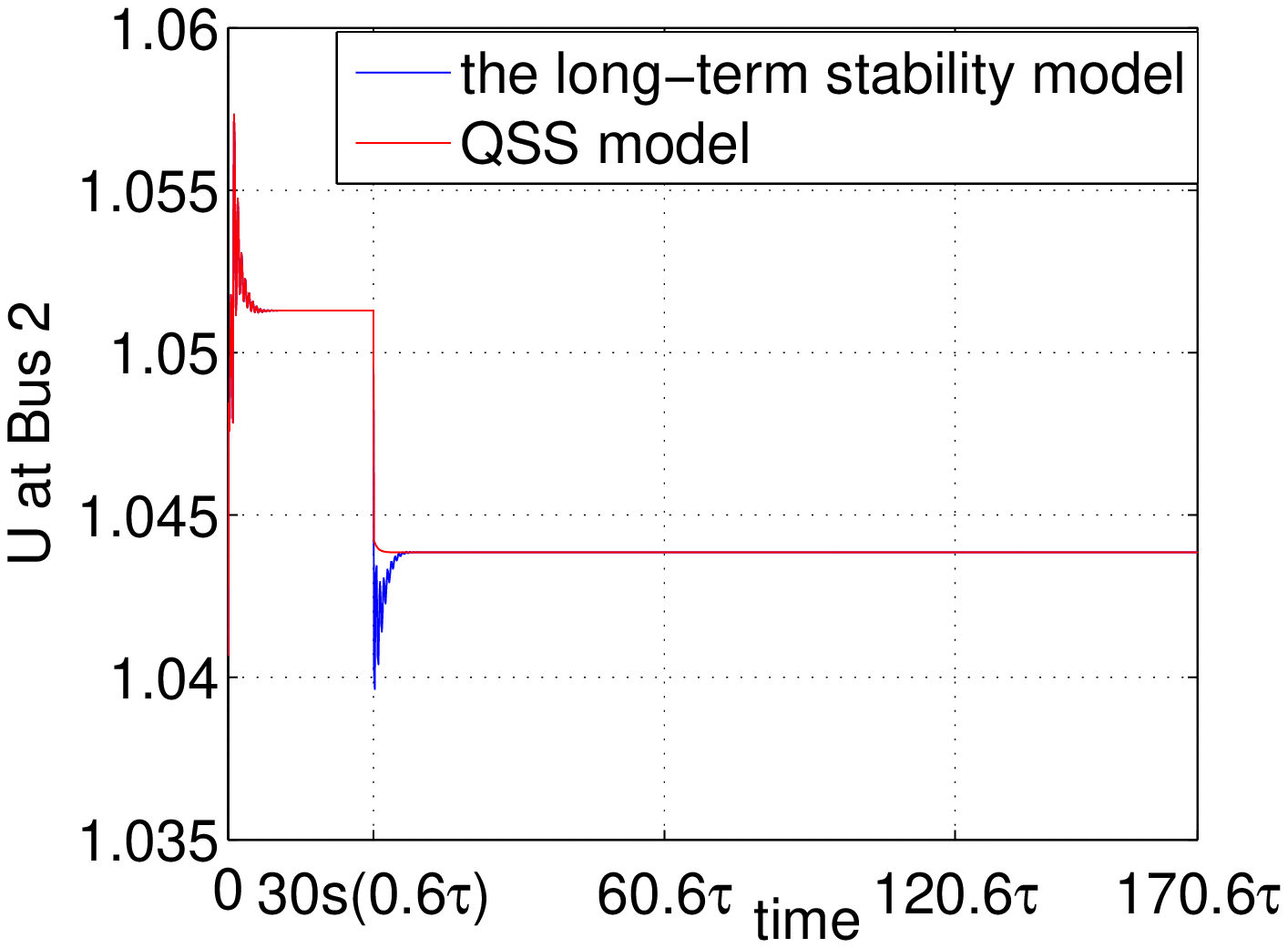}}
\subfloat[]{\label{my14trytransient_31}\includegraphics[width=2.5in ,keepaspectratio=true,angle=0]{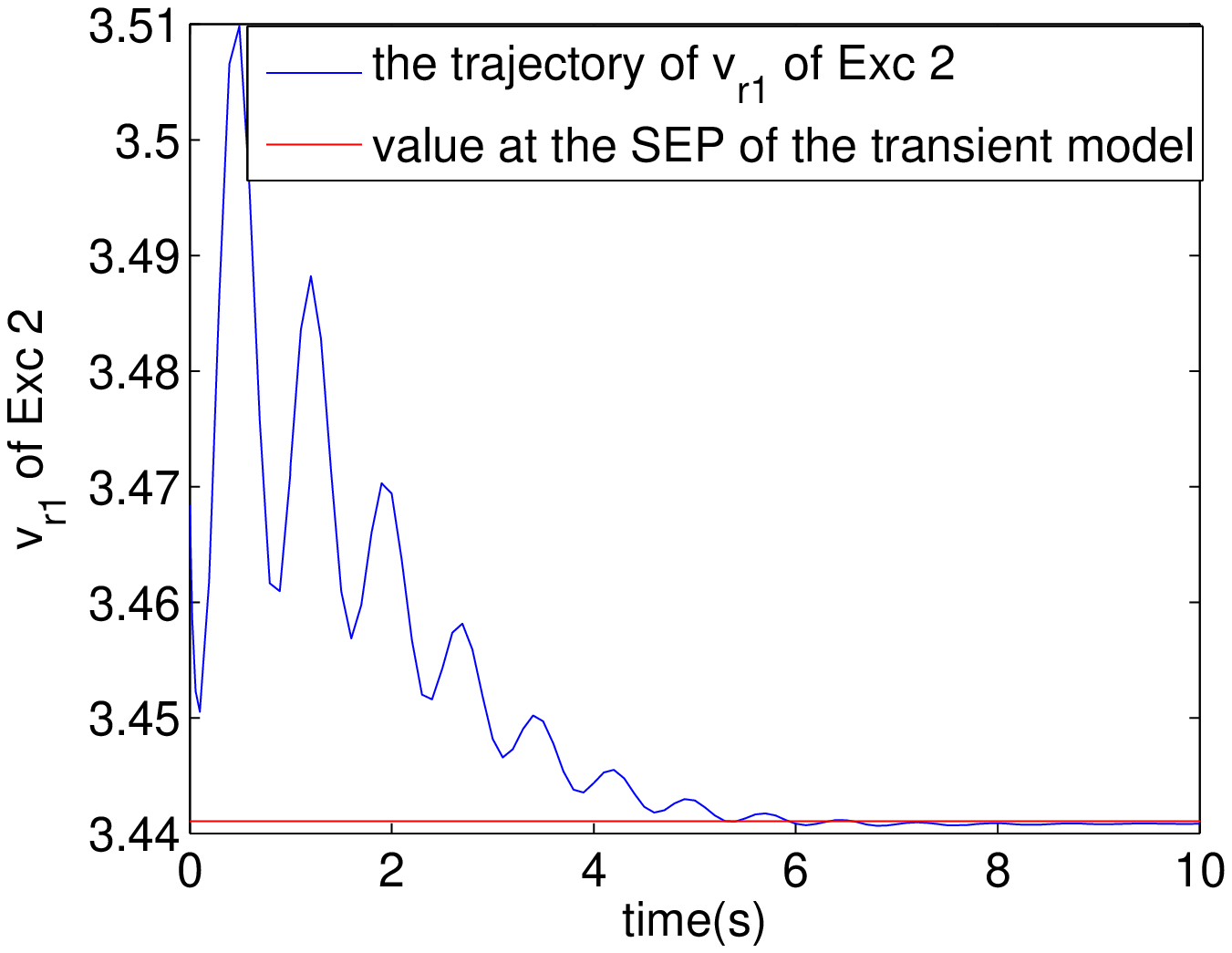}}
\caption{(a). The trajectory comparison of the long-term stability model and the QSS model when LTCs didn't change after 30s. Both models converged to the same long-term SEP. (b). Trajectory of the transient stability model when LTCs changed at 30s. The trajectories starting from $(z_c^1,z_d(1),\bar{x}^1,\bar{y}^1)$ converged to the SEP of the transient stability model.}
\end{figure}

\begin{figure}[!ht]
\centering
\subfloat[]{\label{vfvr1_sepchangetransient_30}\includegraphics[width=2.5in ,keepaspectratio=true,angle=0]{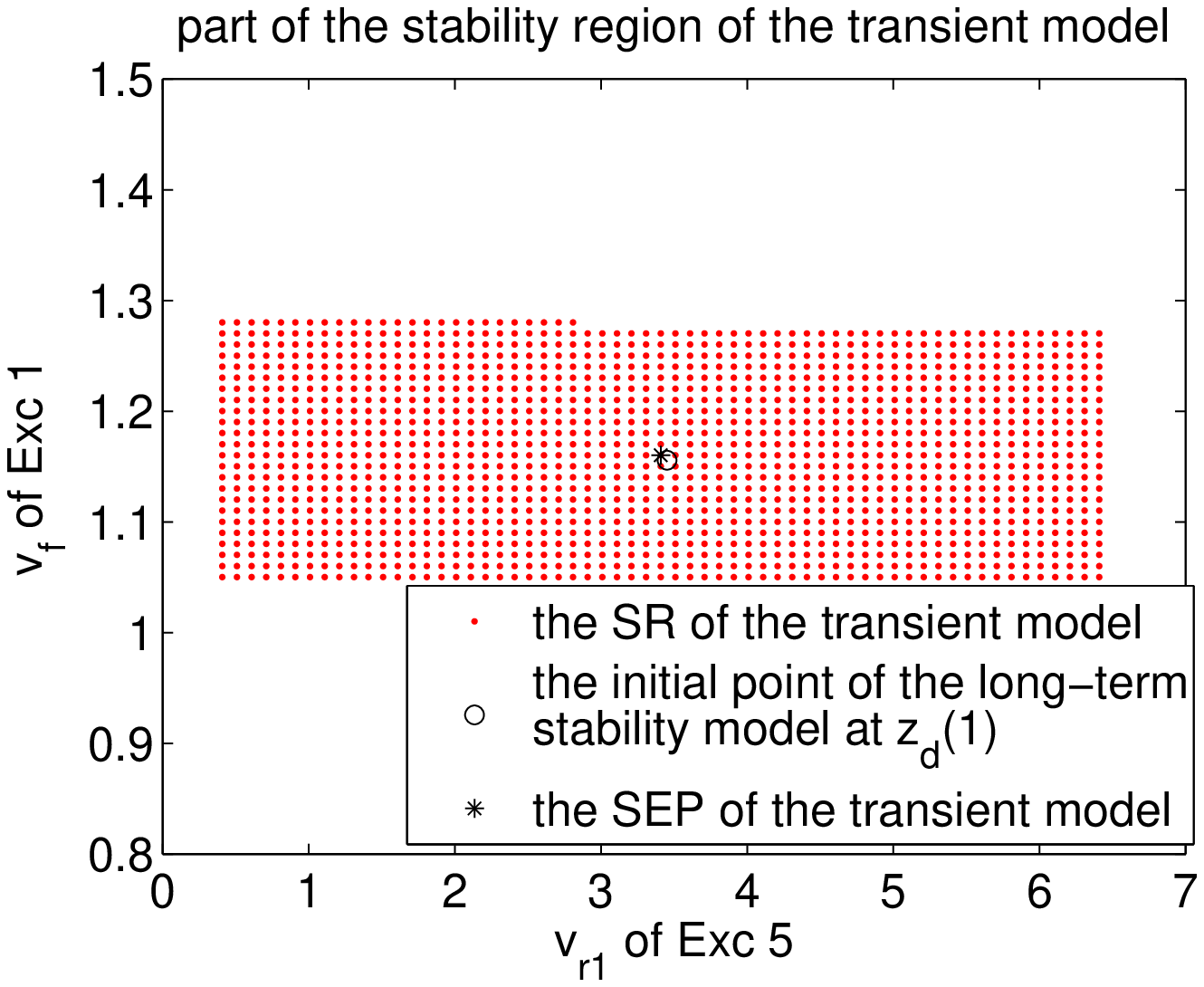}}
\subfloat[]{\label{vfvr1_sepchangetransient_40}\includegraphics[width=2.5in ,keepaspectratio=true,angle=0]{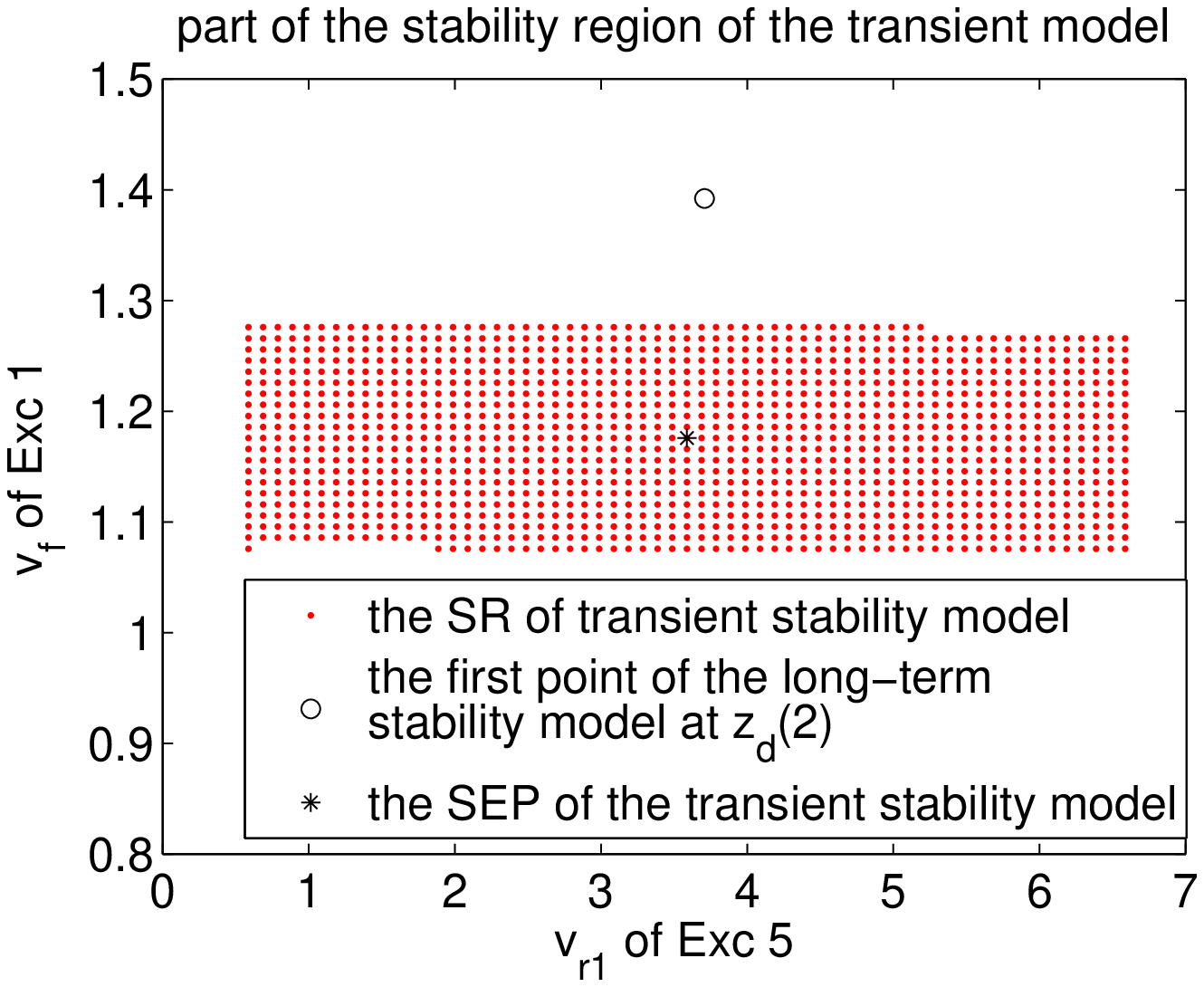}}
\caption{(a). A subspace of the stability region of the transient stability model when $z_d=z_d(1)$. $(z_c^1,z_d(1),\bar{x}^1,\bar{y}^1)$ was inside $A_t(z_c^1,z_d(1),x_{ts}^1,y_{ts}^1)$; (b). The same subspace of the stability region of the transient stability model as (a) when $z_d=z_d(2)$. $(z_c^2,z_d(2),\bar{x}^2,\bar{y}^2)$ was outside $A_t(z_c^2,z_d(2),x_{ts}^2,y_{ts}^2)$.}\label{SR}
\end{figure}

When $z_d$ firstly changed from $z_d(0)$ to $z_d(1)$ at 30s, the long-term stability model jumped to $(z_c^1,z_d(1),\bar{x}^1,\bar{y}^1)$, while the QSS model jumped to $(z_c^1,z_d(1),x_{ts}^1,y_{ts}^1)$. If $z_d$ were frozen from then on, both the long-term stability model and the QSS model converged to the same long-term SEP which is shown in Fig. \ref{my14sepchange_31}. 
From the aspect of stability region, the point $(z_c^1,z_d(1),\bar{x}^1,\bar{y}^1)$ on $\phi_l(\tau,z_{c0},z_{d0},x_0^l,y_0^l)$ was inside the stability region $A_t(z_c^1,z_d(1),x_{ts}^1,y_{ts}^1)$ of the corresponding transient stability model:
\begin{eqnarray}
\dot{x}&=&{f}({z_c,z_d(1),x,y})\qquad\quad x(t_0)={x}_{ts}^1\\\
{0}&=&{g}({z_c,z_d(1),x,y})\nonumber
\end{eqnarray}
which can be seen from Fig. \ref{my14trytransient_31} where trajectory of a fast variables in the transient stability model is plotted. Moreover, Fig. \ref{vfvr1_sepchangetransient_30} shows the intersection of the stability region of the transient stability model in the subspace of two fast variables and we can see that $(z_c^1,z_d(1),\bar{x}^1,\bar{y}^1)$ was inside $A_t(z_c^1,z_d(1),x_{ts}^1,y_{ts}^1)$.

\begin{figure}[!ht]%
\centering
\subfloat[]{\label{my14sepchange_41}\includegraphics[width=2.5in ,keepaspectratio=true,angle=0]{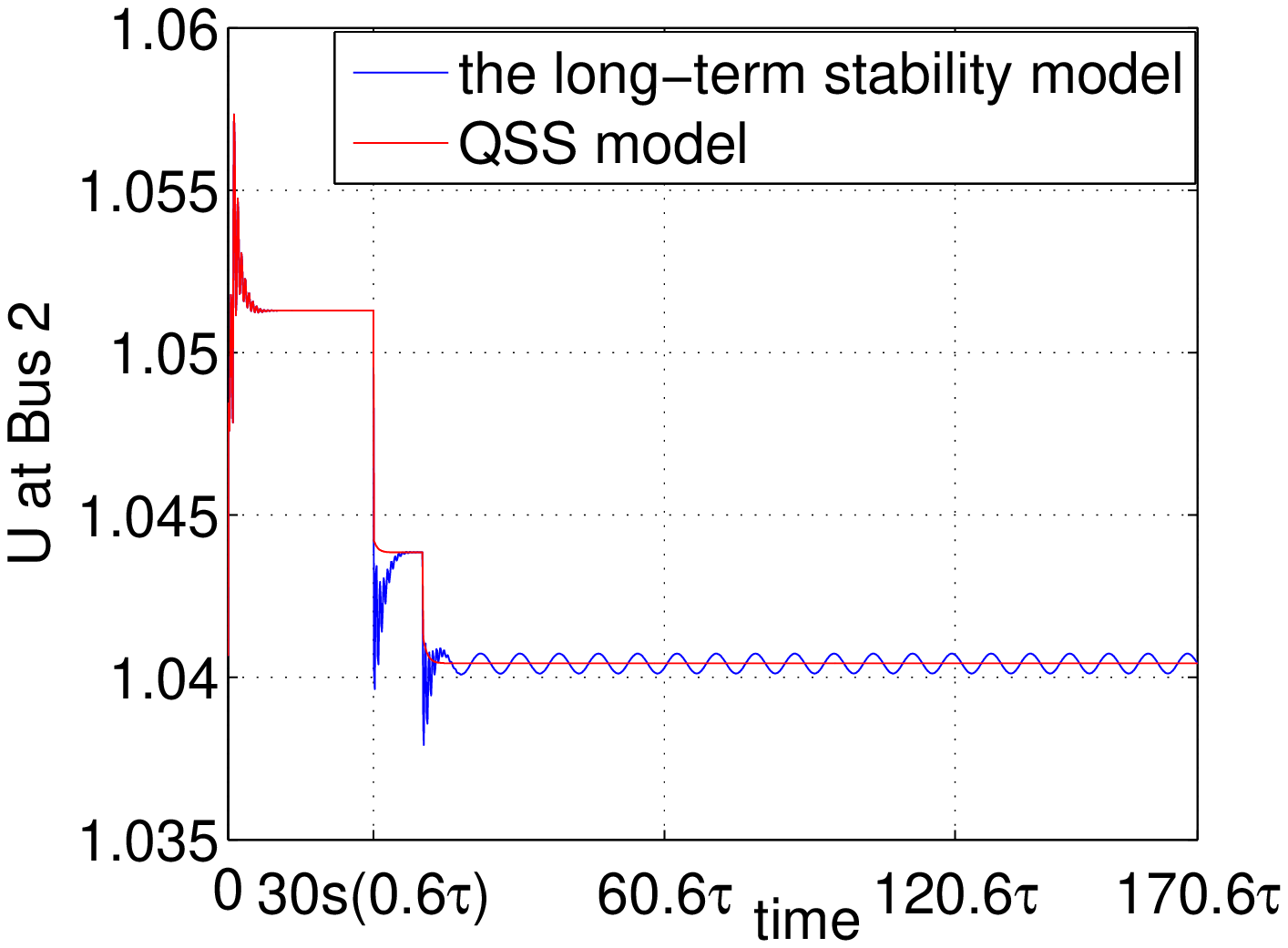}}
\subfloat[]{\label{my14trytransient_41}\includegraphics[width=2.5in ,keepaspectratio=true,angle=0]{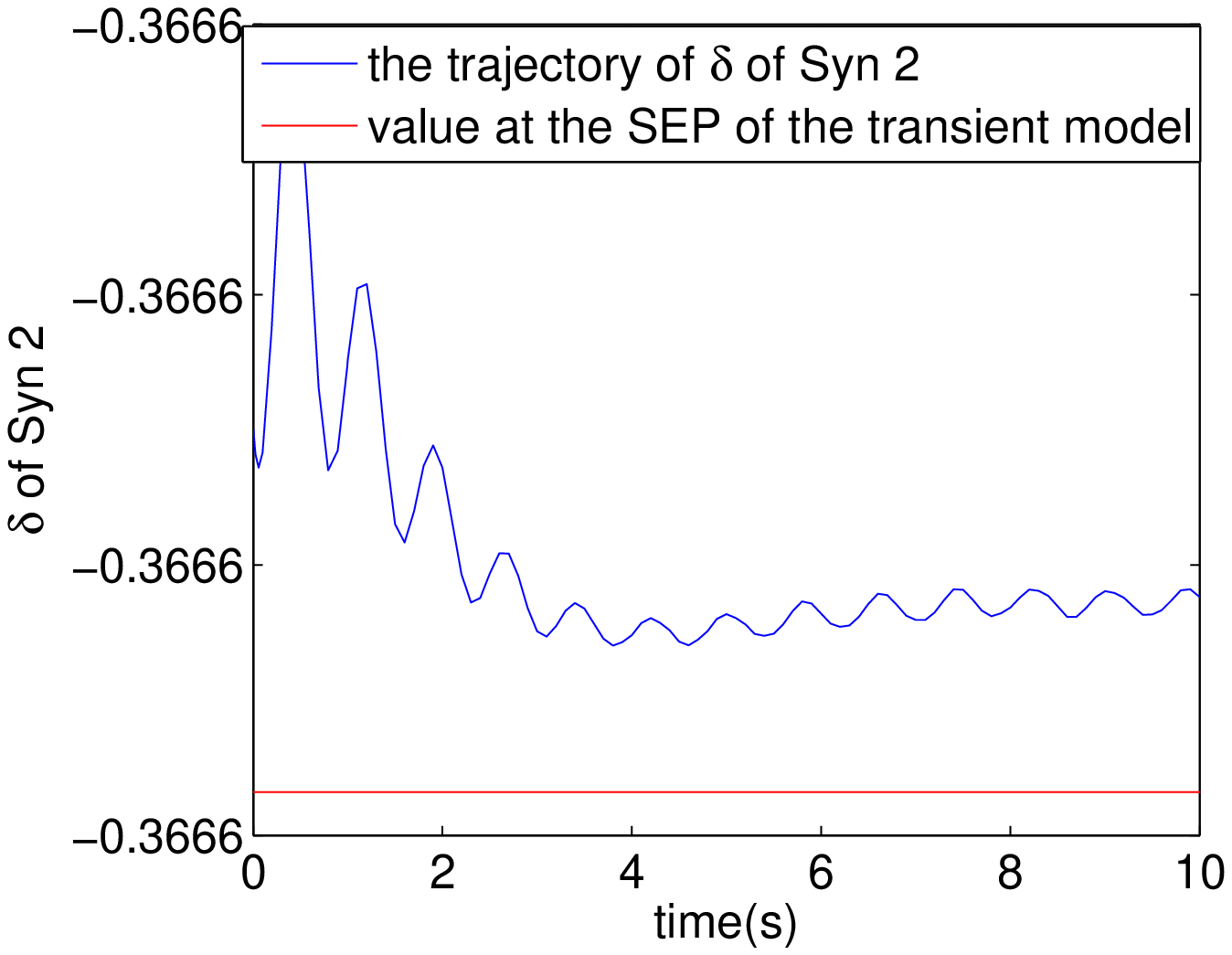}}
\caption{(a). Trajectory comparisons of the long-term stability model and the QSS model when LTCs did change after at 40s. The QSS model didn't capture the dynamics of fast variables and converged to a long-term SEP. (b). Trajectory of the transient stability model when LTCs changed at 40s. The trajectory starting from $(z_c^2,z_d(2),\bar{x}^2,\bar{y}^2)$ didn't converge to the SEP of the transient stability model.}
\end{figure}

However when $z_d$ changed from $z_d(1)$ to $z_d(2)$ at 40s, the long-term stability model was no longer stable which can be seen from the trajectory comparison of the long-term stability model and the QSS model in Fig. \ref{my14sepchange_41}. The trajectory of a fast variables in the corresponding transient stability model is plotted in Fig. \ref{my14trytransient_41}. Additionally, Fig. \ref{vfvr1_sepchangetransient_40} shows the intersection of the stability region of the transient stability model in the same subspace as Fig. \ref{vfvr1_sepchangetransient_30}. We can see that the point $(z_c^2, z_d(2), \bar{x}^2,\bar{y}^2)$ of the long-term stability model
was outside the stability region $A_t(z_c^2,z_d(2),x_{ts}^2,y_{ts}^2)$ of the corresponding transient stability model.

\subsection{{the Numerical Example III:145-bus system}}\label{145bus}
\begin{figure}[!ht]
\begin{minipage}[t]{0.5\linewidth}
\includegraphics[width=2.5in,keepaspectratio=true,angle=0]{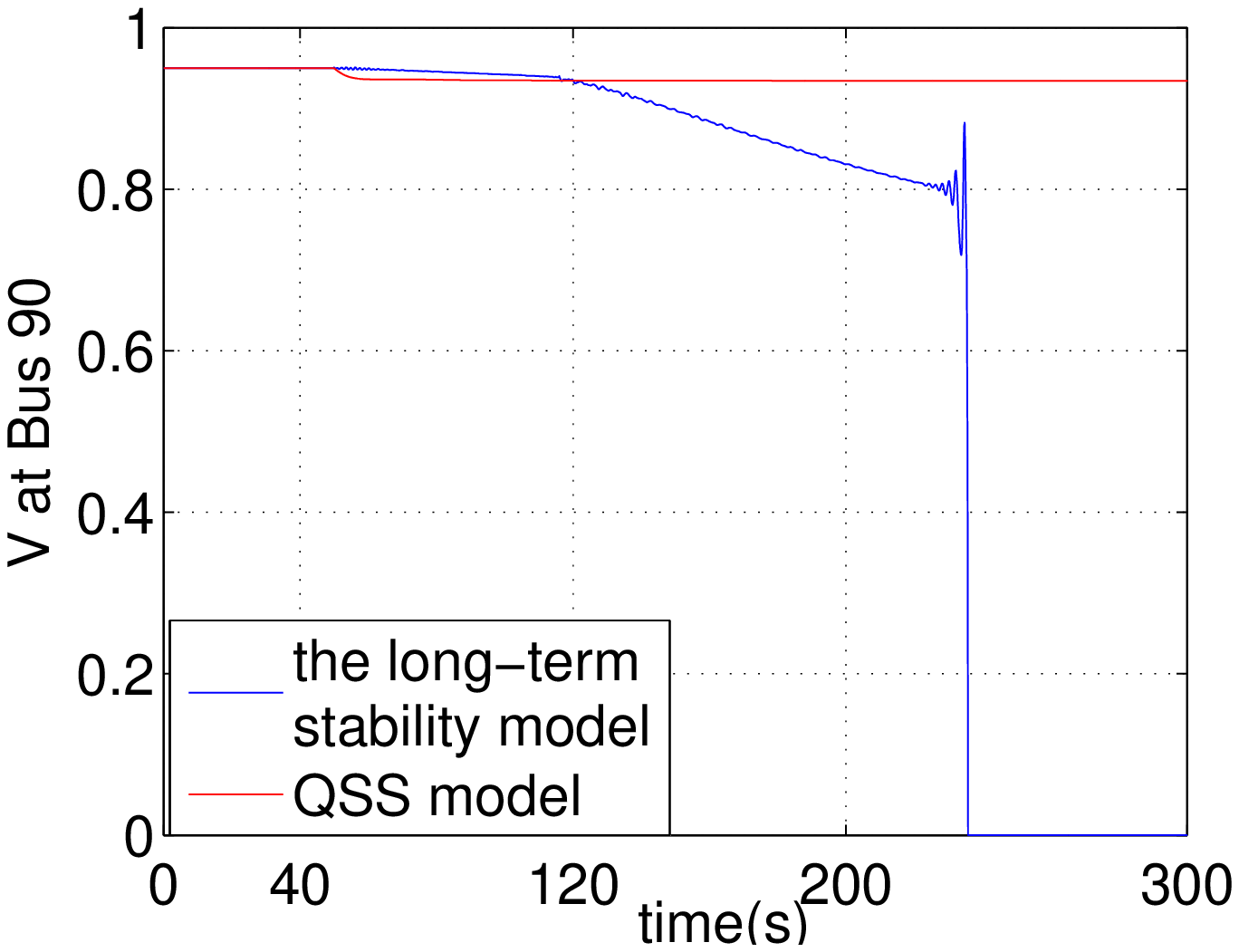}
\end{minipage}%
\begin{minipage}[t]{0.5\linewidth}
\includegraphics[width=2.5in,keepaspectratio=true,angle=0]{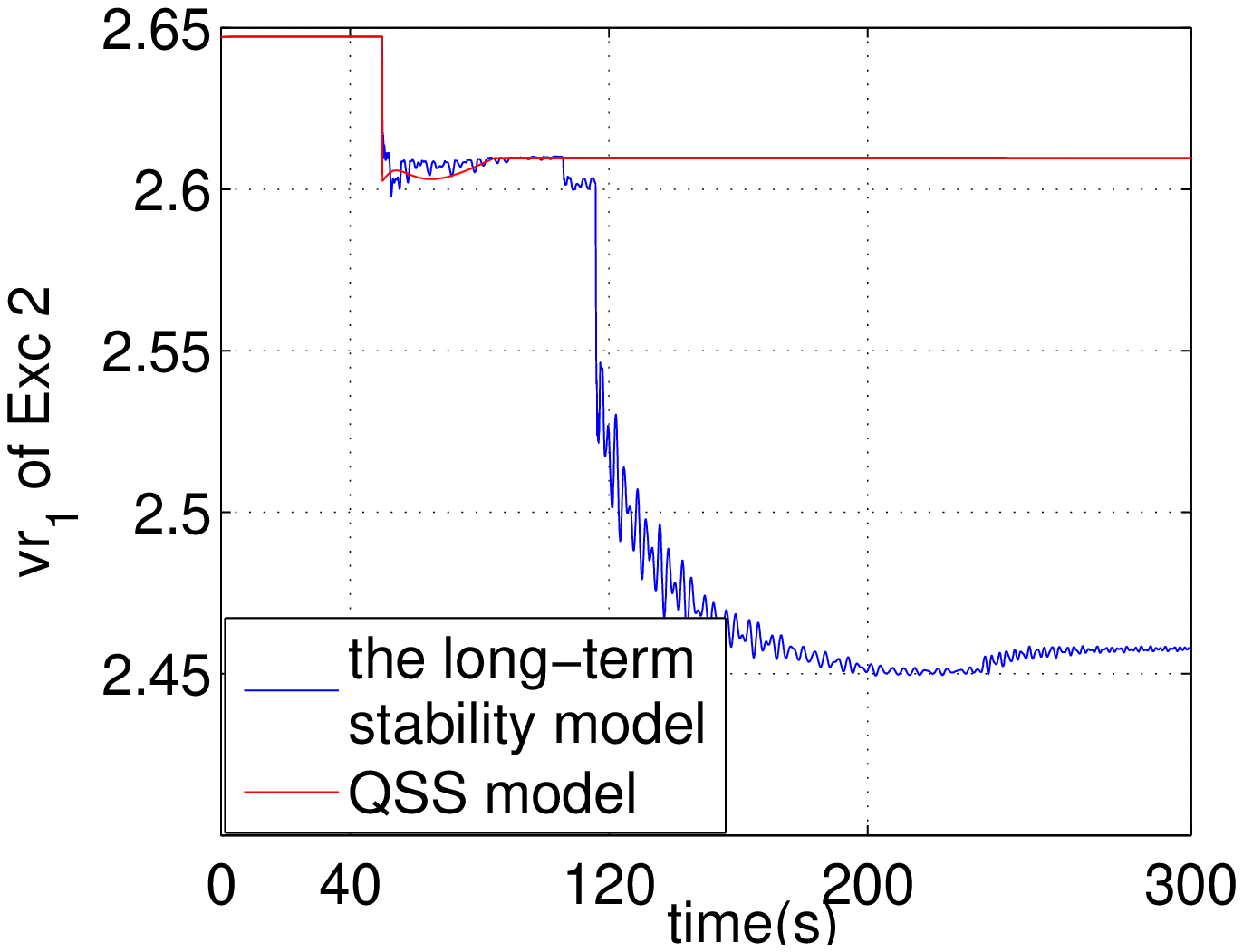}
\end{minipage}
\caption{Trajectory comparisons of the long-term stability model and the QSS model for two variables. In this case, the QSS model converged to a long-term SEP while the long-term stability model suffered from voltage collapse. The QSS model gave incorrect approximations of the long-term stability model.}\label{my145}
\end{figure}

In this IEEE 145-bus system \cite{testcasearchive}\cite{Vittal:article}, comparisons of trajectories in the long-term stability model and the QSS model are shown in Fig. \ref{my145}. In this system, each generator was controlled by AVR and PSS, and there were turbine governors from Generator 10 to Generator 20 and OXLs from Generator 1 to Generator 6. The voltage stability region of the post-fault system was almost the same as that of the pre-fault system.

In this numerical example, when $z_d$ changed at 50s, the trajectory of the QSS model $\phi_q(\tau,z_{c0},z_{d0},x_0^q,y_0^q)$ jumped away from $\Gamma_s$ on which each point was a SEP of the corresponding transient stability model. In other words, the slow manifold of the QSS model was no longer stable. Actually, after $z_d$ jumped at 50s, the equilibrium point of the transient stability model was a type-2 unstable equilibrium point (UEP) which meant the there were two eigenvalues $\lambda$ of $(D_x f\nonumber-D_y f{D_y g}^{-1}D_x g)$ whose real parts were greater than $0$. As a result, the underlying assumption of the QSS model that transient SEP existed and was stable was violated, thus the QSS model failed to correctly approximate the long-term stability model.

\section{\MakeUppercase{Conclusion and Prospectives}}\label{conclusion}
In this paper, we have presented three examples showing that the QSS model can give incorrect approximations of the long-term stability model and provide incorrect voltage stability assessment.  Moreover, causes for failure of the QSS model in those examples are explained and further analyzed in nonlinear system framework. Additionally, we show that sufficient conditions for the QSS model to hold correct approximations should include the following two nonlinear conditions:

\noindent 1). each point on the trajectory of the long-term stability model must be inside the stability region of the corresponding transient stability model;

\noindent 2). the slow manifold of the QSS model must be constrained on the stable component of the constraint manifold.

The failure of the QSS model suggests a necessity to develop a theoretical foundations for the QSS model to ensure correct approximations. Hopefully, an improved QSS model can be developed shortly.

\section{\MakeUppercase{ List of Symbols and Abbreviations }}
\begin{tabular}{ll}
$z_c$ & the vector of long-term continuous state variables\\
$z_d$ & the vector of long-term discrete variable\\
$x$   & the vector of short-term state variables\\
$y$   & the vector of short-term algebraic variables\\
$\ee$ & reciprocal of the maximum time constant among devices\\
$\phi$& trajectory of the dynamical system\\
$A$& stability region of the dynamical system\\
$\Gamma$&constraint manifold of the DAE system\\
$S$&singularity of the DAE system\\
\end{tabular}

\begin{tabular}{ll}
DAE& Differential-Algebraic Equation\\
QSS&Quasi Steady-State \\
HVDC&High-Voltage, Direct Current\\
SVC&Static Var Compensator\\
TG & Turbine Governor\\
AVR& Automatic Voltage Regulator\\
PSS& Power System Stabilizer\\
LTC& Load Tap Changer\\
OXL& Over Excitation Limiter\\
SEP& Stable Equilibrium Point\\
\end{tabular}

\appendix
\section{Parameters of Systems in the Paper}
All test systems are based on models in psat-2.1.6 and parameters for different systems are listed as below.
\subsection{9-bus system}
The system is based on 9-bus test system in psat-2.1.6. There is an exponential recovery load at Bus 5 and there are turbine governors and over excitation limiters for each generator. The contingency is a line-loss of the line between Bus 6-4.
The parameters of the exponential recovery load are:
active power percentage $k_p=40\%$;
reactive power percentage $k_q=40\%$;
active power time constant $T_p=1s$;
reactive power time constant $T_q=1s$;
static active power exponent $\alpha_s=1$;
dynamic active power exponent $\alpha_t=10$;
static reactive power exponent $\beta_s=2$;
dynamic reactive power exponent $\beta_t=20$.

The parameters of the type 1 type turbine governors are:
reference speed $\omega^{0}_{ref}=1$ p.u.;
Droop $R=0.02$ p.u.;
maximum turbine output $p^{max}=2$ p.u. for Generator 1-2 and $p^{max}=4.8$ p.u. for Generator 3;
minimum turbine output $p^{min}=0.3$ p.u.;
governor time constant $T_s=0.1s$;
servo time constant $T_c=0.45s$;
transient gain time constant $T_3=0s$;
power fraction time constant $T_4=12s$;
reheat time constant $T_5=50s$.

The parameters of the over excitation limiters are:
integration time constant $T_0=10s$ for Generator 1-2 and $T_0=30s$ for Generator 3;
maximum field current $i_f^{lim}=2.02$ p.u.;
maximum output signal $v_{oxl}^{max}=100$ p.u.

The parameters of the load tap changers are:
the reference voltage $v_0=1.015$ p.u.;
half of the deadband $d=0.015$ p.u.;
tap step $r=0.05$;
upper tap limit $r^{max}=1.00$;
lower tap limit $r^{min}=0.03$;
the initial time delay $\triangle{T_0}=30s$;
the sequential time delay $\triangle{T_k}=10s$.

\subsection{14-bus system}
The system is based on 14-bus test system in psat-2.1.6. There is an exponential recovery load at Bus 5 and there are turbine governors and over excitation limiters for each generator. The contingency are line losses of the lines between Bus 6-13, Bus 7-9 and Bus 6-11.

The parameters of the type 1 type turbine governors are:
reference speed $\omega^{0}_{ref}=1$ p.u.;
Droop $R=0.02$ p.u.;
maximum turbine output $p^{max}=2$ p.u.;
minimum turbine output $p^{min}=0.3$ p.u.;
governor time constant $T_s=0.1s$;
servo time constant $T_c=0.45s$;
transient gain time constant $T_3=0s$;
power fraction time constant $T_4=12s$;
reheat time constant $T_5=50s$.

The parameters of the load tap changer are:
the reference voltage $v_0=1.005$ p.u.;
half of the deadband $d=0.025$ p.u.;
tap step $r=0.02$;
upper tap limit $r^{max}=1.03$;
lower tap limit $r^{min}=0.98$.
the initial time delay $\triangle{T_0}=60s$;
the sequential time delay $\triangle{T_k}=10s$;

The parameters of the exponential recovery loads are:
active power percentage $k_p=100\%$;
reactive power percentage $k_q=100\%$;
active power time constant $T_p=1s$;
reactive power time constant $T_q=1s$;
static active power exponent $\alpha_s=1$;
dynamic active power exponent $\alpha_t=1.5$ for the load at Bus 9 and $\alpha_t=5$ for the loads at Bus 10 and 14;
static reactive power exponent $\beta_s=2$;
dynamic reactive power exponent $\beta_t=2.5$ for the load at Bus 9 and $\beta_t=10$ for the loads at Bus 10 and 14.

The parameters of the over excitation limiter are:
integration time constant $T_0=30s$ for Generator 3 and $T_0=10s$ for the others;
maximum field current $i_f^{lim}=1.22$ p.u. for Generator 3 and $i_f^{lim}=0$ p.u.;
maximum output signal $v_{oxl}^{max}=100$ p.u.
\renewcommand{\refname}{REFERENCES}

\end{document}